# What is Unusual about the Third Largest Geomagnetic Storm of Solar Cycle 24?

**N. Gopalswamy[1], S. Yashiro[1,2], S. Akiyama[1,2], H. Xie[1,2], P. Mäkelä[1,2], M.-C. Fok[1], C. P. Ferradas[1,2]**

[1]Heliophysics, NASA Goddard Space Flight Center, Greenbelt, MD.

[2]Department of Physics, The Catholic University of America, Washington DC.

Corresponding author: Nat Gopalswamy (nat.gopalswamy@nasa.gov)

**Key Points:**

- Coronal mass ejection characterized by prolonged acceleration, rotation, and high-density content results in the intense geomagnetic storm

- Simulations with a kinetic code confirm that the ring current injection involves both solar wind dynamic pressure and electric field

- Empirical formulas for predicting Dst based on solar wind electric field work when the magnetic cloud has no high density-enhancement

## Abstract

We report on the solar and interplanetary (IP) causes of the third largest geomagnetic storm (2018 August 26) in solar cycle 24. The underlying coronal mass ejection (CME) originating from a quiescent filament region becomes a 440 km/s magnetic cloud (MC) at 1 au after ~5 days. The prolonged CME acceleration (for ~24 hrs) coincides with the time profiles of the post-eruption arcade intensity and reconnected flux. Chen et al. (2019) obtain lower speed since they assumed that the CME does not accelerate after ~12 hrs. The presence of multiple coronal holes near the filament channel and the high-speed wind from them seem to have the combined effect of producing complex rotation in the corona and IP medium resulting in a high-inclination MC. The Dst time profile in the main phase steepens significantly (rapid increase in storm intensity) coincident with the density increase (prominence material) in the second half of the MC. Simulations using the Comprehensive Inner Magnetosphere-Ionosphere (CIMI) model shows that a higher ring current energy results from larger dynamic pressure in MCs. Furthermore, the Dst index is highly correlated with the main-phase time integral of the ring current injection that includes density, consistent with the simulations. A complex temporal structure develops in the storm main phase if the underlying MC has a complex density structure during intervals of southward interplanetary magnetic field. We conclude that the high intensity of the storm results from the prolonged CME acceleration, complex rotation, and the high density in the 1-au MC.

## Plain Language Summary

Powerful coronal mass ejections (CMEs) are responsible for very intense geomagnetic storms due to the out of the ecliptic component of the magnetic field in the CME flux rope or in the sheath if shock-driving (Gosling 1993). The 2018 August 26 storm was very intense, but the CME was inconspicuous and weak near the Sun. However, over an extended period of time the CME accelerated slowly and picked up adequate speed to cause an intense storm. Due to the presence of coronal holes near the eruption region, the CME rotated in such a way that the CME magnetic field and Earth's magnetic field can efficiently couple to transfer energy into the magnetosphere to cause the geomagnetic storm. The energy transfer is expedited by the presence of dense material deep inside the CME.

## 1 Introduction

It is well established that intense geomagnetic storms with a Dst index $< -150$ nT are almost always associated with coronal mass ejections (CMEs), while weaker storms with Dst $> -150$ nT can be caused by CMEs and stream interaction regions (SIRs). CMEs causing geomagnetic storms are generally fast and wide indicating they are very energetic (see e.g., Gopalswamy 2018 and references therein). Occasionally, storms are caused by slower CMEs as observed in the coronagraph field of view (FOV) (Zhang et al. 2007). Many of these CMEs continue to accelerate beyond the coronagraph FOV and attain sufficient speed to drive shocks at large distances from the Sun that can be detected in situ or via purely kilometric type II radio bursts (Gopalswamy 2006; Gopalswamy et al. 2010). During the weak solar cycle 24, the frequency and intensity of geomagnetic storms is unusually low (Gopalswamy 2012; Richardson 2013; Kakad et al. 2019). Towards the end of this cycle, an intense storm has been observed on 2018 August 26 with a Dst index of –175 nT. Only two storms in solar cycle 24 are stronger than this event: the 2015 March 17 (–222 nT) and 2015 June 23 (–204 nT) storms (see e.g., Liu et al. 2015; Gopalswamy et al. 2015a; Wu et al. 2016; Webb and Nitta 2017). The 2018 August 26 event is characterized by weak solar eruption, significant flux rope rotation in the corona and interplanetary medium, and an intense geomagnetic storm, as first reported by Chen et al. (2019). These authors identify the solar source of this storm as a filament channel eruption and track the CME propagation in the corona and interplanetary medium. The event has also been reported to have widespread space weather effects at Earth (Zakharenkova et al. 2021; Abunin et al. 2020) and Mars (Thampi et al. 2021). Although some authors claim that this is a stealth CME (Mishra and Srivastava 2019; Piersanti et al. 2020; Nitta et al. 2021), the near-surface signatures are clear although weak (Chen et al. 2019; Abunin et al. 2020).

One of the key findings by Chen et al. (2019) is that the CME flux rope rotated between the Sun and Earth, resulting in a unipolar magnetic cloud (MC) with its axis pointing to the south throughout the cloud. The high intensity of the geomagnetic storm has been attributed to the long-duration southward field possibly enhanced by the compression due to following high speed stream (HSS) (Chen et al. 2019; Abunin et al. 2020; Nitta et al. 2021). It is well known that the storm strength represented by the Dst index has a high correlation with the solar wind electric field VBz, where V is the speed and Bz is the out of the ecliptic component of the interplanetary magnetic structure such as a MC (Murayama 1982; Gonzalez and Tsurutani 1987; Wu and Lepping 2002; Wang et al. 2003a; Kane 2005; Gonzalez and Echer 2005; Gopalswamy et al. 2008; 2015b). The 2018 August 26 event seems to be an exception because the observed maximum value of –VBz is too small to account for the storm intensity of –175 nT. Using Bz = –16.4 nT and V = 400 km/s from Chen et al. (2019), the resulting –VBz has a maximum of 6560 km/s nT (in GSE coordinates). Using the empirical formula for the minimum value of Dst given by (Gopalswamy et al. 2008),

Dst = –0.01 VBz – 32 nT                    (1)

 we can get a maximum strength of only – 98 nT, about half of the observed –175 nT. Therefore, how the weak eruption and the resulting MC caused the third largest storm is a mystery. The 1-au speed of the CME used by Chen et al. (2019) is underestimated because (i) not fully tracking the CME acceleration and (ii) using a slightly different initial boundary of the interplanetary CME (ICME). The actual leading edge has a speed of 440 km s$^{-1}$ (see later). Even if we use V = 440 km s$^{-1}$ in Equation 1, we get Dst = - 104 nT.

One of the factors not considered in the above works is the density within the MC. The solar wind density has been considered as a factor in determining the geoeffectiveness of interplanetary structures (see e.g., Weigel 2010 and references therein). A high solar wind density can lead to higher density in the magnetospheric plasma sheet (Borovsky et al. 1998), and the latter can influence the ring current amplitude (Jordanova et al. 2003). MHD simulations show that increased solar wind density during intervals of southward Bz can increase the bow-shock compression ratio resulting in increased magnetospheric energy dissipation rate (Lopez et al. 2004). Towards predicting Dst, Murayama (1982) was the first to consider the effect of the solar wind dynamic pressure ($P_f$) by incorporating it into the ring current injection ($Q \sim VBz \times P_f^{1/3}$). Several variants of Q have been considered by Gonzalez et al. (1989). Fenrich and Luhmann (1998) consider density enhancement inside MCs like we do in this paper. Wang et al. (2003b) consider both Q and the decay rate as functions of solar wind electric field Ey = VBz and $P_f$. They find that $P_f$ (which is proportional to the solar wind density) can increase the ring current injection during Bz <0 and decrease the ring current decay time during Bz >0. Using such an injection term, Xie et al. (2008) find that the Dst peak value of a storm increases when there is a large enhancement in $P_f$ during the main phase of a storm.  Using a Q similar to that of Wang et al. (2003b),  Le et al. (2020) show that the storm strength defined by the SYM-H index is highly correlated with the time-integral of the injection over the main phase (see also Zhao et al. 2021). Weigel (2010) shows that the solar wind electric field results in a larger magnetospheric response when the solar wind density is higher. Based on the above discussion, we conclude that it is worth examining the effect of the high density inside the MC to see if it can explain the observed Dst peak value and its time profile in the main phase of the 2018 August 26 storm.

In Section 2, we summarize the observations from the Sun to 1 au. In Section 3 we analyze the observations and present new results regarding CME kinematics, MC structure, and Dst time profile. In Section 4, we discuss the results and provide a summary of the investigation in section 5.

## 2 Observations

The primary objective of this paper is to provide a physical description of the solar and interplanetary circumstances that led to the intense 2018 August 26 geomagnetic storm. The provisional Dst index obtained from the Kyoto World Data Center (WDC) for Geomagnetism (http://wdc.kugi.kyoto-u.ac.jp/dstdir/, Nose et al. 2015) shows that the Dst index attains a minimum value of –175 nT.  The source of the 2018 August 26 storm is a MC associated with a filament channel eruption on 2018 August 20 that results in a white-light CME. The filament channel, the post eruption arcade (PEA), and coronal dimming are observed at several wavelengths by the Atmospheric Imaging Assembly (AIA, Lemen et al. 2012) on board the Solar Dynamics Observatory (SDO). The AIA images are also used in identifying the coronal holes near the filament channel. The filament channel is along the neutral line of a large-scale magnetic region identified in the Helioseismic and Magnetic Imager (HMI, Scherrer et al. 2012). H-alpha images obtained by the Big Bear Solar Observatory are used to identify the filament and the filament channel (http://www.bbso.njit.edu/Research/FDHA/menu.html). The white-light CME is observed by the Large Angle Spectrometric Coronagraph (LASCO, Brueckner et al. 1995) on board the Solar and Heliospheric Observatory (SOHO) and the Sun Earth Connection Coronal and Heliospheric Investigation (SECCHI, Howard et al. 2008) on board the Solar Terrestrial Relations Observatory (STEREO). The combined SOHO and STEREO images help us track the CME from the Sun to Earth. We use OMNI data (https://omniweb.gsfc.nasa.gov/) to describe the plasma and magnetic properties of the interplanetary CME (ICME).

Figure 1 provides an overview of the eruption region (filament channel) with nearby coronal holes in an SDO/AIA 193 Å image taken several hours before the eruption. The filament channel extends from N50W10 to N10W40. The centroid is roughly at N20W10, close to the disk center. A dark filament is present at the northern end of the filament channel. Two coronal holes are located on the east (CH-E) and west (CH-W) sides of the filament channel.  There is another large coronal hole (CH-S) to the south of the eruption region, probably connected to CH-W.

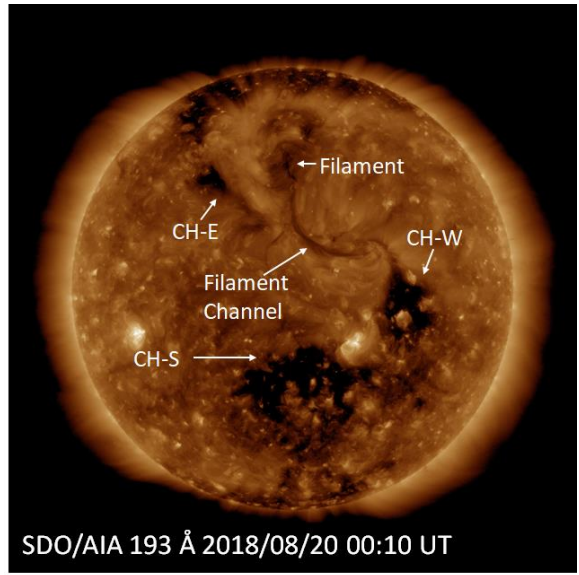

**Figure 1.** An overview of the source region and its surroundings of the 2018 August 20 coronal mass ejection: a filament channel oriented in the NE-SW direction. Only a small section of the filament channel contains a filament as marked. Coronal holes located on the east and west side of the channel are marked as CH-E and CH-W, respectively. There is also another coronal hole to the south, marked as CH-S. The SDO/AIA 193 Å image was taken at 00:10 UT, several hours before the onset of the eruption.

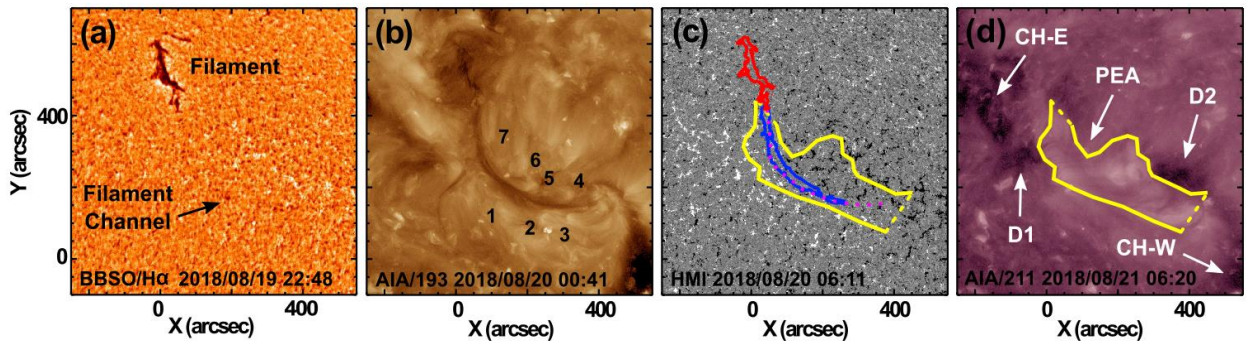

**Figure 2.** (a) H-alpha image of the source region before eruption showing the filament fragment in the north and tiny fragments along the filament channel. (b) SDO/AIA 193 Å image showing coronal cells numbered from 1 to 7 on either side of the filament channel. (c) SDO/HMI line of sight magnetogram at 06:11 UT with an outline of the H-alpha filament (red), AIA 193 Å filament channel (blue), and the H-alpha trace of the filament channel (pink dots) marked. Also superposed is the foot-points of the post eruption arcade (yellow lines) extracted from the SDO/AIA 211 Å image taken at 06:20 UT on 2018 August 21 (d). In (d), the two coronal holes (CH-E and CH-W) are marked along with the core dimming regions D1 and D2 located just outside the PEA.

## 3 Analysis and results

Figure 2 shows more details of the source region from various sources. The northeast end of the filament channel has a clear filament, and the rest of the channel has tiny filament fragments as can be seen in the H-alpha image (Fig. 2a). The filament can also be seen in the SDO/AIA 193 Å image (Fig. 2b). The HMI magnetogram shows that the filament channel is located along the polarity inversion line (PIL) of a large-scale bipolar magnetic region (Fig. 2c). The east and west side of the PIL have positive and negative polarities, respectively. Tadpole-shaped coronal cells line up on either side of the filament channel, seven of them marked in Fig. 2b.  The coronal cells 1-3 are located on the positive side of the PIL while cells 4-7 are located on the negative side. The cells are similar to the chromospheric fibrils with the head of the tadpoles located on a majority-polarity magnetic element (Martin 1998; Sheeley et al. 2013). The field direction in the cell is the same as that of the filament channel, so we infer from Fig. 2c that the field direction is southward along the filament channel. The helicity sign is negative (left-handed) because the azimuthal field above the filament channel goes from east to west, in agreement with the hemispheric rule. The filament channel eruption is marked by the formation of a PEA starting around 08:00 UT on 20 August 2018 that takes about a day to reach its full size.  The outline of the PEA (enclosed by the yellow lines) is overlaid in Fig. 2c,d. The eruption of the filament channel is accompanied by core dimmings (D1, D2) located on either side of the polarity inversion line (Fig. 2d). The line connecting the D1 and D2 has a tilt of ~ − 6⁰, which is smaller than the tilt of the PEA (~ −30⁰) and the PIL (~ −45⁰).

## 3.1 CME kinematics

The white-light CME is listed in the SOHO/LASCO CME catalog (https://cdaw.gsfc.nasa.gov, Yashiro et al. 2004; Gopalswamy et al. 2009a) as a slowly accelerating CME (~5.4 m s$^{-2}$) with a first appearance time of 21:24:05 UT on 2018 August 20. The linear sky-plane speed is 126 km/s, which is expected to be much smaller than the true three-dimensional (3D) speed because of the severe projection effects in a disk-center eruption. At the time of the eruption onset, STEREO Ahead (STA) was located at E108 from Earth. Therefore, in STA view, the eruption is slightly behind the west limb, so the speed measured from STA is closer to the 3D speed. Although extremely faint, the eruption can be seen at 08:30 UT in STA/COR1 image, see: (https://stereo ssc.nascom.nasa.gov/browse/2018/08/20/ahead_20180820_cor1_rdiff_512.mpg). The CME first appears in the STA/COR2 FOV around 12:00 UT. We use coronal images from SOHO and STA to fit a graduated cylindrical shell (GCS, Thernisien 2011) flux rope. Snapshots of the CME from SOHO and STEREO are shown in Fig. 3 along with the GCS flux rope overlaid on the coronagraph images. In addition to SOHO/LASCO and SECCHI/COR2 images, we have included SECCHI's Heliospheric Imager (HI) data in the GCS fit. The flux rope leading edge is at a height of ~64 Rs in the HI1A image shown Fig. 3d,h taken at 04:49 UT on August 22. The tilt of the GCS flux rope axis turns out to be 12⁰, indicating a counterclockwise rotation of ~18⁰ with respect to the line connecting the dimming regions.

 One of the interesting features in these images is the core of the CME, which has a brightness similar to that of the leading edge early on but becomes the dominant feature later on (in the HI1 FOV). This is the vertical feature in the middle of the FOV in Fig. 3d. This feature is also observed in the HI2 FOV and in-situ when the flux rope arrives at Earth.

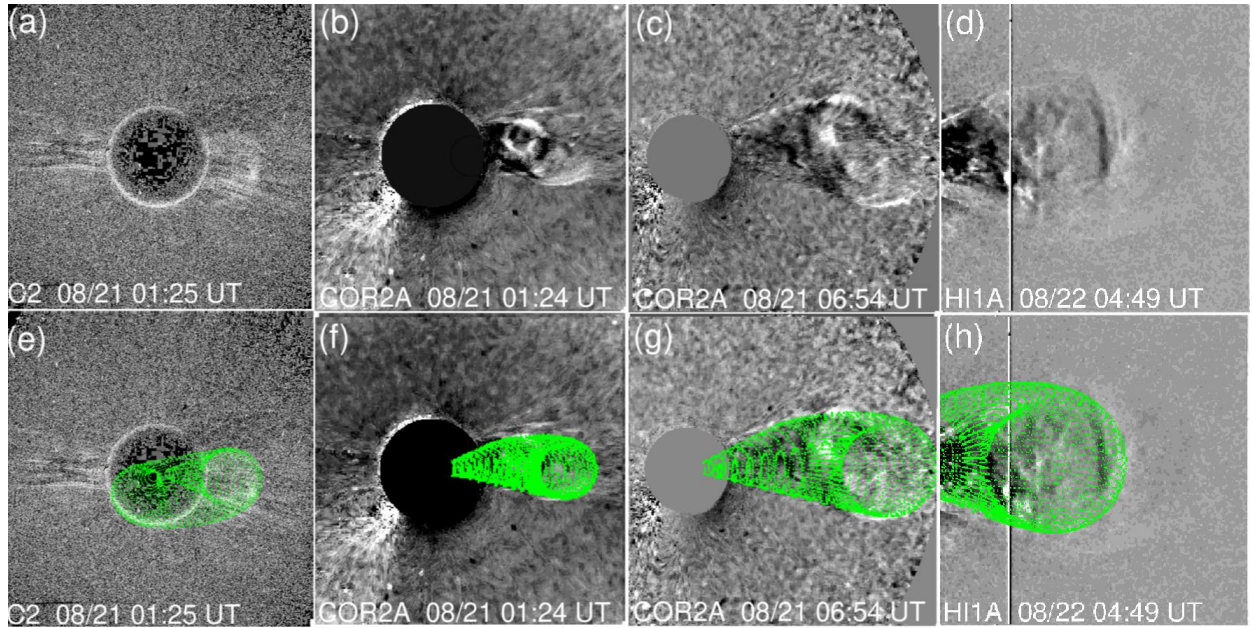

**Figure 3.** Snapshots of the CME in question at our times: (a) 2018 August 21 at 01:25 UT (LASCO C2), (b) at 01:24 UT (SECCHI COR2A), (c) at 06:54 UT (SECCHI COR2A), and (d) at 04:49 UT on August 22 (SECCHI HIA). The corresponding flux ropes fitted to the CME are shown the bottom panels (e-h). The leading edge of the flux rope is at 64.3 Rs in the HI1A FOV. We track the leading edge of the flux rope until it reaches ~133.5 Rs in HI2A FOV at 13:31 UT on August 23. Beyond this distance, the features are too faint to make measurements. However, playing HI2 movies, we can see the CME disturbances blowing past Earth around midday on August 25. As expected, the 3D speed is ~400 km/s within the LASCO FOV, which is much higher than the sky-plane speed (~126 km/s). The average acceleration within the LASCO FOV is ~7.5 m s$^{-2}$.

## 3.2 Signatures of Magnetic Reconnection

Figure 4 shows the time evolution of the PEA intensity (I), its time derivative (dI/dt), and the intensity in the dimming regions in comparison with the height-time plot of the GCS flux rope's leading edge. Although the PEA is well defined, its intensity is extremely weak, so no enhancement is observed in the GOES soft X-ray light curve. The situation is similar to the source regions of polar CMEs, whose kinematics can be understood using the EUV intensity of the PEA and its time derivative (Gopalswamy et al. 2015c). This is because both PEA and the CME flux rope are created by magnetic reconnection and the PEA intensity closely follows the CME speed (Zhang et al. 2001). dI/dt mimics the Neupert effect (Neupert 1968; Dennis & Zarro 1993) and hence follows the CME acceleration. In the 2018 August 20 event, dI/dt becomes positive at the same time as the dimming onset around 10:00 UT and drops to zero level around 22:00 UT the next day (see Fig. 4b). There are several bumps in dI/dt. The CME acceleration from the leading edge of the GCS flux rope corresponds to the third and largest bump in dI/dt. Both the CME acceleration and dI/dt drop to low values around 06:00 UT on August 21 remaining positive until about 22 UT. The close correspondence between dI/dt and CME acceleration is remarkable given the weakness of the PEA. The cumulative reconnected (RC) flux ($\Phi_r$) reaches a steady value of ~ $(1.6 \pm 0.19) \times 10^{21}$ Mx around 08:00 UT on August 21. The instantaneous RC flux computed every 2 hours ($d\Phi_r/dt$) shows a time variation very similar to those in dI/dt and CME acceleration. The low values of dI/dt, CME acceleration, and the RC flux are clear between 08:00 and 22:00 UT on August 21. The clear dip around 21 UT on August 20 is also simultaneous in $d\Phi_r/dt$ and dI/dt. The first broad bump in $d\Phi_r/dt$ has a counterpart in dI/dt, but the latter has a double structure, which probably is not seen in $d\Phi_r/dt$ due to the low time resolution employed. The height-time plot in Fig. 4b shows that CME continues to accelerate into the HI1A FOV, reaching ~50 Rs by the time the acceleration ceases around 22 UT on August 21.

The acceleration seems to be powered by the reconnection the whole time. Evidence for the continued increase of CME speed beyond ~100 Rs due to the continued effect of magnetic reconnection in the source region has been presented by Temmer et al. (2011). Sachdeva et al. (2015) have also shown that the evolution of slow CMEs is not affected by the drag force below the range 15–50 Rs. Here we have shown direct evidence from the evolution of PEA arcade, RC flux, and CME acceleration that the propelling force can act at distances >50 Rs. Chen et al. (2019) assumed that the acceleration ceases when the CME leading edge is at a height of ~17 Rs. However, they also reported continued gradual separation of flare ribbons for 24 hrs. The continued ribbon separation is consistent with the gradual increase in PEA brightness in Fig. 4b because the flare ribbons correspond to the feet of the PEA. The ribbon separation indicates continued reconnection evidenced by the increase in $\Phi_r$ (see Fig. 4c). The continued reconnection implies that the propelling force has not ceased, consistent with the positive acceleration of the CME shown in Fig. 4b.

Slowly accelerating CMEs are generally associated with filament eruptions outside active regions and can cause type III bursts, type II bursts, and large SEP events if they accelerate to high enough speeds (Kahler et al. 1986; Gopalswamy et al. 2015d; Liu et al. 2016; Cliver et al. 2019). Some slowly accelerating CMEs can become superalfvenic at distances of tens of solar radii to drive a shock and produce purely kilometric type II radio bursts (Gopalswamy 2006). In some cases, the shock may not cause a type II bursts (a radio-quiet shock) but a weak shock is observed in the solar wind data (Gopalswamy et al. 2010). Examination of ground-based and space-based radio observations shows that the 2018 August 20 eruption is not associated with any radio emission.

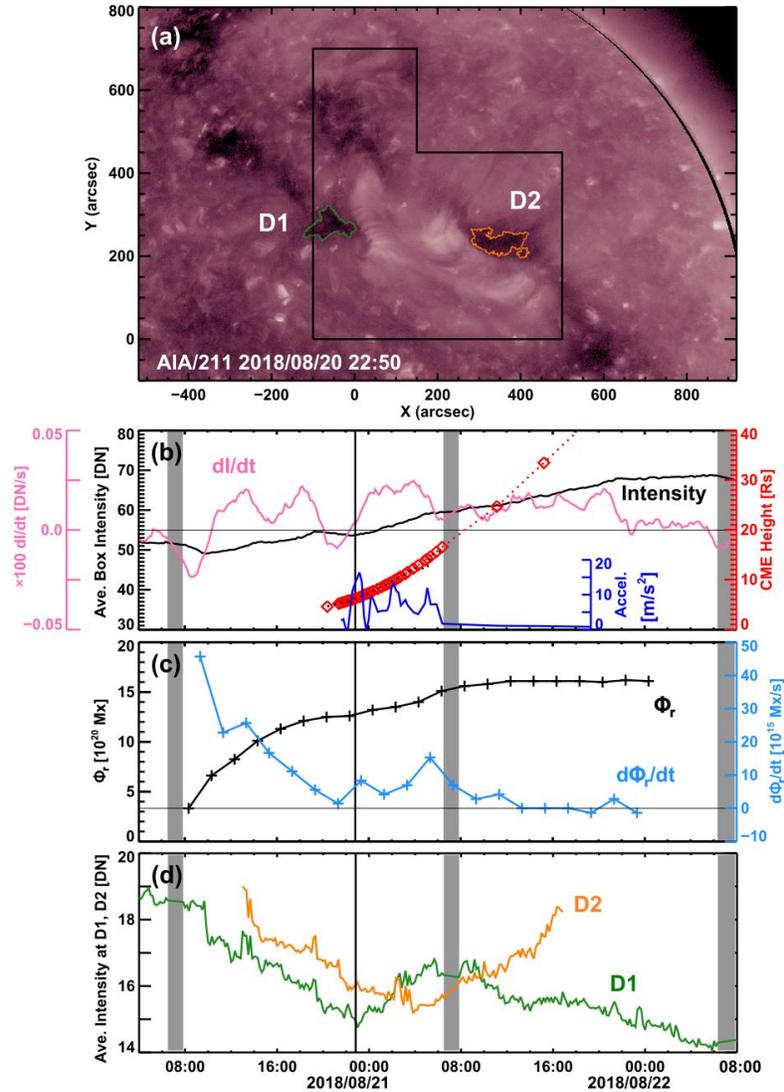

**Figure 4.** (a) SDO/AIA 211 Å image showing the PEA and dimming regions D1 (green contour) and D2 (orange contour). The box encloses the area where the PEA is contained. (b) The average EUV intensity (I, black curve) within the box in (a) and its time derivative (dI/dt, pink curve) plotted as a function of time. The area corresponding to the dimming regions is excluded in computing the average intensity in units of data number (DN). The leading-edge height of the GCS flux rope (red diamonds) along with the quadratic fit (dotted line) to the height-time data points. The last two data points correspond to the HI1A FOV. The CME acceleration derived from the height-time measurements is shown in blue. (c) The time evolution of the RC Flux ($\Phi_r$) and its time derivative $d\Phi_r/dt$ computed from PEA every two hours. (d) The average EUV intensity in the diming regions D1 (green curve) and D2 (orange curve). The three gray vertical bands denote intervals of SDO data gap. The vertical dark line marks the time of the SDO/AIA image in (a).

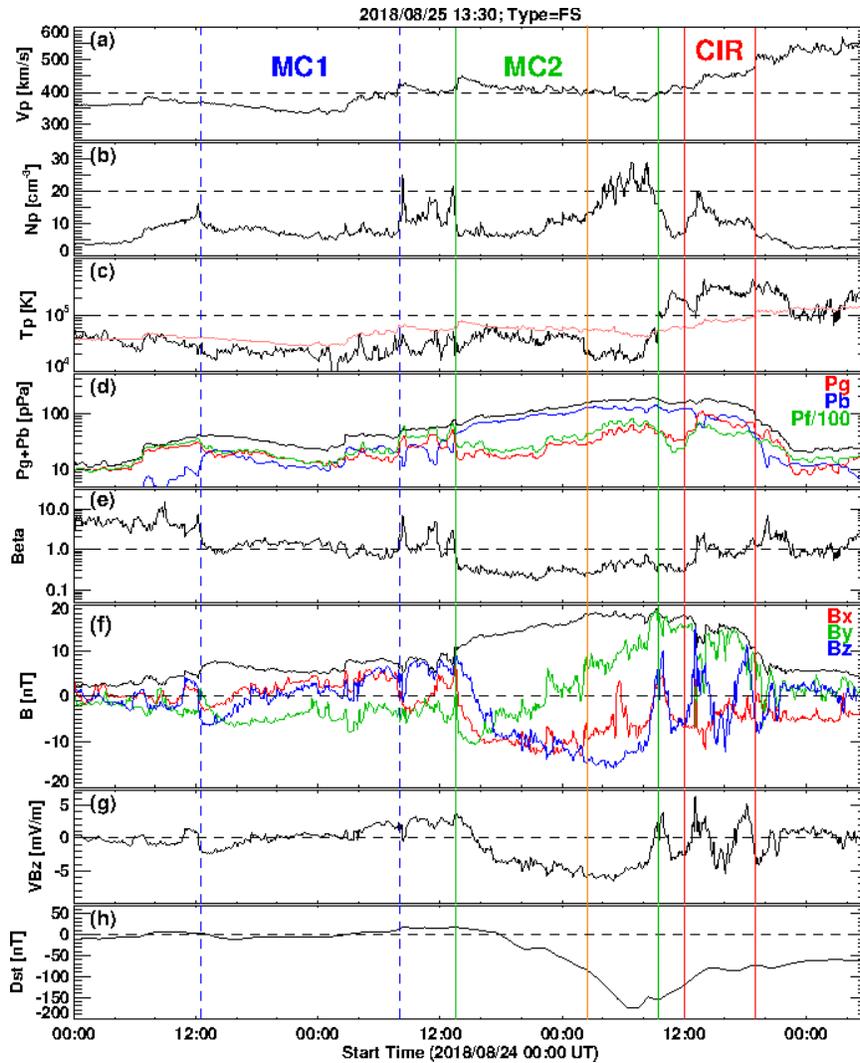

**Figure 5.** Solar wind observations from OMNI for the period 2018 August 24 – 27. (a) Solar wind speed ($V_p$), (b) proton density ($N_p$), (c) proton temperature ($T_p$) along with the expected temperature (orange line), (d) gas ($P_g$ – red curve), magnetic ($P_b$ – blue curve), and flow ($P_f$ – green curve) pressures and the total pressure ($P_g$+$P_b$ – black curve), (e) plasma beta, (f) total magnetic field strength (B) along with the three components $B_x$ (red curve), $B_y$ (green curve), and $B_z$ (blue curve) in GSE coordinates, (g) solar wind electric field (solar wind speed times the $B_z$ component of the magnetic field), (h) the $D_{st}$ index showing the intense geomagnetic storm with a slope change in the main phase at the instance marked by the vertical orange line (02:30 UT on August 26). This line also marks the start of the density increase that lasts until the rear boundary of the MC. The $D_{st}$ data are from the World Data Center, Kyoto. The vertical green lines mark the boundaries of the magnetic cloud based on $T_p$ (beginning and end of MC interval), beta (beginning of MC interval), and B (beginning and end of MC interval). The vertical blue dashed lines mark the boundary of a preceding MC on August 24 (MC1). The MC on August 25 (MC2) has its $B_z$ negative throughout and hence designated as fully southward (FS) MC meaning it is a high-inclination MC with its axial field pointing southward. The By component rotates from west to east, so this is a left-handed (WSE MC). MC2 was followed by a CIR interval indicated by the vertical red lines. The ambient solar wind ahead of MC1 has a speed of ~350 km/s as can be seen at the beginning of the plot (before 4:00 UT on August 24).

### 3.3 Interplanetary CME and the geomagnetic storm

The interplanetary counterpart of the CME is a MC according to the criteria of Burlaga et al. (1981): flux rope structures with enhanced magnetic field, smooth rotation of the azimuthal component, and low proton temperature and/or plasma beta.  The MC arrives Sun-Earth L1 at 13:00 UT on August 25 and lasts until ~09:00 UT on August 26 (labeled MC2 in Fig. 5).  Unlike the proton temperature signature, the plasma beta signature is well defined, so we use it to define the first boundary of the MC. Chen et al. (2019) used the temperature signature to identify the initial MC boundary to be a couple of hours after our boundary. MC2 is preceded by another MC labeled MC1 in Fig. 5. Both MC1 and MC2 are slow and are not driving shocks. However, there is some compressed plasma separating the MCs that arrives at ~08:00 UT on August 25. In shock driving MCs a well-defined sheath is expected (Yue and Zong 2011; Manchester et al. 2014). The MC intervals are marked by the vertical blue and green lines. MC1 is a bipolar MC (smooth rotation of the Bz component) satisfying the MC criteria (low proton temperature, but the plasma beta hovers around 1). On the contrary, MC2 is a unipolar MC (smooth rotation of the By component). The central speed of MC2 is ~400 km/s with a slightly higher (440 km/s) and lower (370 km/s) speeds at the leading and trailing edges, respectively. This indicates that the MC continues to expand at 1 au.  The MC2 leading edge speed is much larger than the ambient solar wind speed of ~350 km/s as can be seen in Fig. 5a before 4:00 UT on August 24. MC2 is immediately followed by a corotating interaction region (CIR), with a stream interface around 13:00 UT on August 26. The CIR is identified based on the increase in density, temperature, and magnetic field strength during the positive gradient of the solar wind speed (Wilcox and Ness 1965; Belcher and Davis 1971; Gosling et al. 1972; Smith and Wolfe 1976; Barnes and Simpson, 1976). Hereafter, we denote MC2 by MC and do not discuss MC1. The magnetic field strength in the MC has a peak value of 18 nT. The high inclination MC has a WSE configuration (negative helicity). The Bz component reaches a peak value of about –15 nT. A flux rope fit to the in-situ data using the Lepping et al. (1990) method confirms the negative helicity and high inclination of the MC with a radius of 0.13 au.

The feature that stands out in the solar wind plots in Fig. 5b is the proton density that starts increasing at ~22:00 UT on August 25, attains a peak value between 20 and 30 $cm^{-3}$. There are 5 large peaks with density >20 $cm^{-3}$, the last two reaching ~30 $cm^{-3}$. The density drops to ~5 $cm^{-3}$ just after the MC rear boundary. The high-density region is also the coolest part of the MC. The opposite trends in density and temperature resulted in a gas pressure that only slightly increases in the region. The magnetic pressure is much larger, so it dominates in the total pressure, which smoothly increases from the beginning of the MC and drops only after the end of the cloud interval. Corresponding to the increase in the gas pressure is the increase in plasma beta but beta stays below 1. We discuss the origin of the high-density material in Section 4.5.

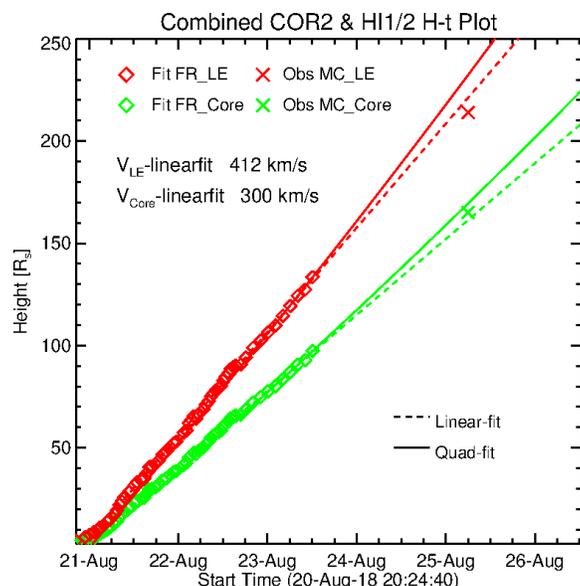

**Figure 6.** The height-time history of the CME flux rope leading edge (red data points) and its core (green data points). Linear and quadratic fits to the height-time data points are shown. The linear fit is closer to the in-situ arrival of the MC leading edge (the compressed material arriving at 08:00 UT as noted in Fig. 5).

The extended height-time plot of the CME as tracked in the FOV of COR2, HI1, and HI2, is shown in Fig. 6. The CME attains roughly a constant speed of ~412 km/s after it finishes accelerating around 22 UT on August 21 when the flux-rope LE is at a height of 50 Rs (and the core at 35 Rs (see Fig. 4(b)). The event is also cataloged in the HELCATS list, which gives the speed in the HI1 FOV as ~286 km s$^{-1}$ (https://www.helcats-fp7.eu/catalogues/event_page.html?id=HCME_A__20180821_01). In the HI2 FOV, one can see the CME blowing past Earth on August 25 (see https://stereo-ssc.nascom.nasa.gov/cgi-bin/images). The CME speed (V in km/s) is related to the total RC flux ($\Phi_r$ in $10^{21}$ Mx): V = 298×$(\Phi_r)^{0.79}$ (Gopalswamy et al. 2018). Inserting the observed $\Phi_r$ of $1.6×10^{21}$ Mx, we get V = 423 km/s, which is in good agreement with the speed from the height-time measurements. The linear fit to the height-time data points is in good agreement with the arrival time of the MC disturbance. When the MC disturbance at 1 au, the high-density region is ~50 Rs behind, which is also consistent with the increase in density in the MC. The quadratic fit would imply a 1-au arrival time of 21 UT on August 24, about 11 hrs ahead of what the linear fit indicates. After the acceleration ends, the CME flux rope seems to propagate at constant speed or slightly decelerating since the in-situ data point of the MC disturbance is located slightly below the linear fit curve. The speed implied by the CME kinematics in Fig. 6 matches with the MC leading edge speed of 440 km/s, differing by <4%.

The Dst index in Fig. 5 starts decreasing about 4 hrs after the Bz in the MC starts turning south. The solar wind electric field VBz attains its minimum value of –6520 km/s nT in the high-density interval at ~05:00 UT on August 26, following which the Dst index reaches its minimum value (–175 nT) two hours later. The Dst time profile shows a remarkable slope change starting around 02:00 UT on August 26, at which time the Dst = –85 nT. The slope changes from –12.5 nT/hr to –22.5 nT/hr, which is a steepening by 77%. The time of the slope change coincides precisely with the time of temperature drop and density increase in the MC (and hence with the gas pressure – see Fig. 5b,c,d). Since the speed of the MC does not change much through the MC interval, the five-fold increase in density should increase the dynamic pressure by the same factor. This gives a clue to the possible mechanism that causes the slope change. The steepening Dst profile indicates that the density increase (or the dynamic pressure of the high-density material) seems to have made the MC more geoeffective. In hindsight, such a slope change can be found in the largest storm in solar cycle 23 (2003 November 20) that has a final Dst = –422 nT (the provisional Dst is –472 nT, see Gopalswamy et al. 2005). The underlying MC has high density material, later confirmed to be prominence material (Sharma and Srivastava 2012). However, neither of these works recognizes the coincidence of the density increase with the steepening of the Dst profile. From the final Dst data, we see that the slope changes from –33.5 nT/hr to –83.5 nT (not shown) when the density increase starts. Recently, Cheng et al. (2020) report on an opposite case: when the density drops significantly during the main phase, the storm strength is accordingly reduced.

The unusual Dst profile indicates that the minimum Dst deviates significantly from the one predicted by empirical relations. We have already shown this to be the case in the introduction using Equation 1. Another empirical relation that considers the storm main-phase duration ($\Delta t$ in hr) is (Wang et al. 2003a):

$$\text{Dst} = -19.01 - 8.43\ (-\langle VBz \rangle)^{1.09}(\Delta t)^{0.3} \quad (2)$$

where $\langle VBz \rangle$ is the average over the main phase of the storm in units of mV/m in GSM coordinates. With $\Delta t = 13$ hr and $-\langle VBz \rangle = 4.74$ mV/m, we get Dst = –121 nT, which is slightly better compared to the Dst from Eq. (1), but the observed Dst is still 45% lower. This is not consistent with the suggestion by Chen et al. (2019) that the high intensity of the 2018 August 26 storm is due to the enhanced strength and duration of Bz alone.

In order to illustrate the importance of density, we compare the 2018 August 26 event with another event (2010 May 29) of similar solar wind parameters but has no significant density enhancement (see Fig. 7). The 2010 May 29 storm is due to a high inclination MC with negative helicity and associated with the 2010 May 23 halo CME originating from a filament eruption region centered around N16W10. The source magnetic configuration is very similar to that of the 2018 August 20 CME. The white-light CME has a higher sky-plane speed (258 km/s) than the 2018 August 20 CME (https://cdaw.gsfc.nasa.gov/CME_list/halo/halo.html). Fig. 7 shows that the density inside the MC has an average value of 6 cm$^{-3}$ and there is no significant enhancement in the second half of the MC. In the first half, there is a small enhancement over a 5-hr interval, starting at 00 UT on May 29 that has a peak value of ~10 cm$^{-3}$. There is a large density enhancement outside the back of the MC due to a CIR formed by a high-speed solar wind stream.

The observed minimum Dst value of the 2010 May 29 storm is only –80 nT, less than half of the minimum Dst in the 2018 August 26 storm. Using the observed –VBz = 4979 km/s nT and –$\langle VBz \rangle = 4.14$ mV/m in the empirical formulas (1) and (2), we get the minimum Dst values as –82 nT and –107 nT, respectively (Empirical formula (1) uses GSE coordinates, while (2) uses GSM coordinates). Note that Equation 1 gives a value very close to the observed value (–80 nT), whereas Equation 2 predicts a stronger storm (–107 nT). The two parameters that differ significantly between the two events are the density and the minimum Dst value: higher density results in a stronger geomagnetic storm on 2018 August 26. The strengthening of the storm coincides with the start of the higher-density interval. The empirical formulas for Dst seem to work for events with "normal" densities.

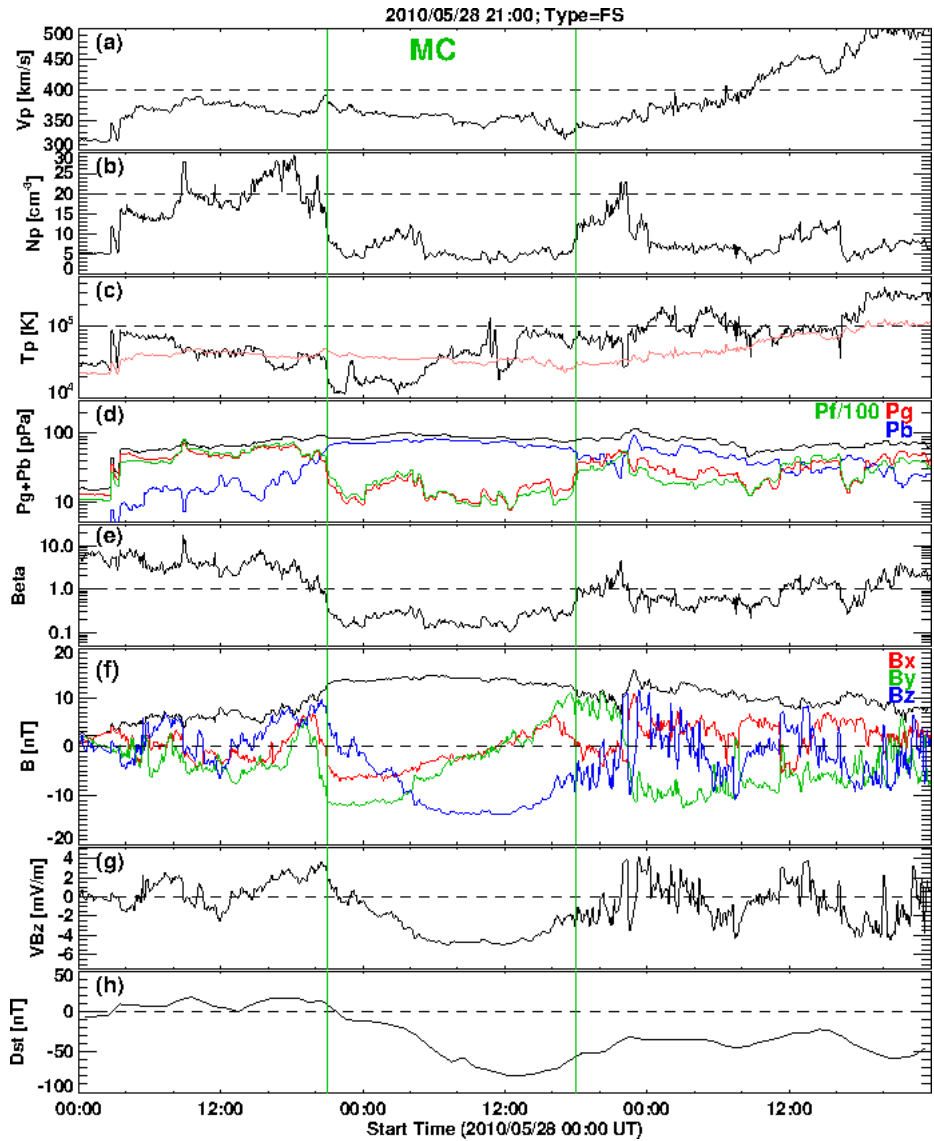

**Figure 7.** Solar wind parameters as in Fig. 5 but for the 2010 May 28 MC that resulted in a moderate storm (−80 nT). The two parameters that look distinctly different from the ones in Fig. 5 are the proton density and the minimum value of the Dst.

The density variation inside some MCs is more structured. The 2014 April 11 MC shown in Fig. 8, is also an FS MC, so Bz <0 throughout the MC interval. The MC has three intervals with different density variations: (i) constant density (~4 cm⁻³) during the first 8 hours of the MC, (ii) slow increase from 4 cm⁻³ to ~7 cm⁻³ over the next 14 hours, and (iii) high density (~18 cm⁻³) in the last 6 hours. During the low-density interval (i), Bz and VBz increase in amplitude but the Dst index hovers slightly above 0 nT. The Dst starts decreasing when the density starts increasing in interval (ii) while Bz and VBz level off. At 22:00 UT on April 11, Dst reaches –34 nT. Further increase in density is accompanied by a slight steepening of the Dst, which reaches a local minimum value of –61 nT at 01:00 UT on April 12. The Dst starts increasing when the Bz magnitude decreases, but the continued increase in density prolongs the storm. Another local minimum in Dst (–83 nT) occurs marking the noticeable decease in Bz magnitude at ~07 UT on April 12. At this time, the density rapidly increases to ~18 cm$^{-3}$ (interval iii) resulting in a Dst of –87 nT, the peak strength of the storm at 10:00 UT on April 12. The peak value of –VBz (3000 km/s nT) when used in Equation 1 yields a Dst of only –62 nT compared to the observed –87 nT. The 40% stronger geomagnetic storm seems to be due to the increased density in the MC. This event illustrates that a combination of density and Bz variations dictate the evolution of the Dst index in the main phase.

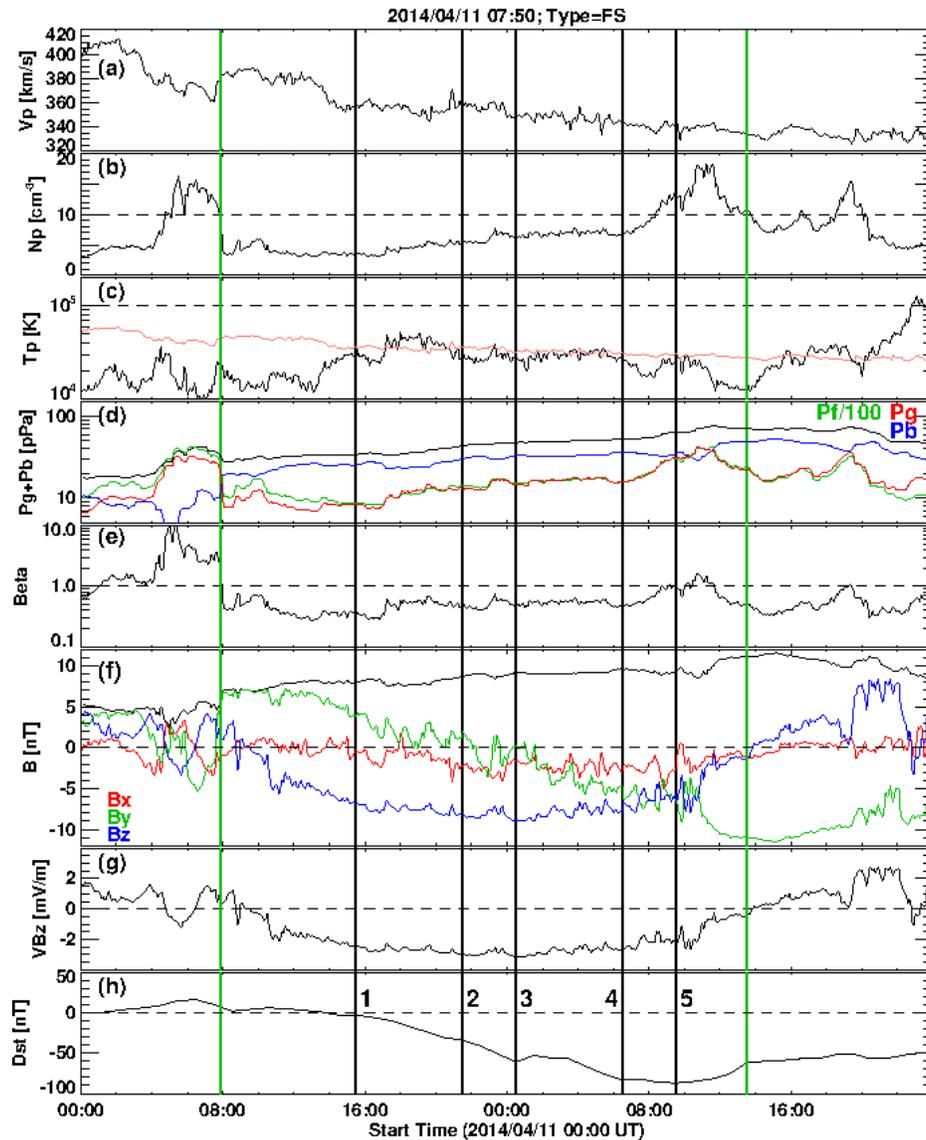

**Figure 8**. Solar wind parameters of the 2014 April 11 MC (between the vertical green lines) and the associated Dst index as in Figs. 5 and 7. The vertical black lines indicate 1. the time when the Dst index started negative excursion; 2. the time of slope change when Np reaches a higher value of ~ 7 cm$^{-3}$. 3. the time of local dip in Dst, corresponding to the upward turning of VBz (decrease in electric field). 4. local Dst minimum followed by a slight recovery. 5. time of Dst minimum. Np peaks when VBz declines significantly. The plasma beta briefly exceeds 1 at this time. Steady recovery of the storm starts at the end of the MC, where VBz = 0.

Table 1 compares the properties of the three MCs discussed above and the associated geomagnetic storms. The MCs are unipolar (FS), of similar size and central speed (Vc), and a slightly longer duration for the 2014 MC. The three MCs differ in proton densities (Np), especially the peak values. From the last three rows in the table, we see that the 2018 storm is much stronger than the other two storms, which are of similar strength (–80 nT and –87 nT). The MCs underlying the 2010 and 2014 storms have similar speeds, but much different Bz. Therefore, |VBz| is higher in the 2010 event. However, the higher |VBz| does not lead to a stronger storm. The main reason is the effect of the density enhancement in the back of the 2014 MC. Comparing the 2018 and 2010 events, we see that the 2018 MC is slightly faster and has a slightly higher |Bz|, so it has a higher |VBz| by ~24% yet it resulted a much stronger storm. In this case also, the higher density in the 2018 MC seems to make the difference.

**Table 1.** Solar wind parameters associated with the geomagnetic storms on 2018 August 26, 2010 May 29 and 2014 April 11 from the OMNI data

| Property | 20180826 | 20100529 | 20140411 | Remark |
|---|---|---|---|---|
| MC type | FS (WSE) | FS (WSE) | FS (ESW) | Unipolar MCs |
| MC duration (hr) | 20 | 21 | 29.7 | |
| MC radius (au) | 0.13 | 0.09 | 0.12 | Lepping et al. Fit |
| Main phase duration (hr) | 12 | 14 | 19 | |
| <Beta> | 0.31 | 0.22 | 0.52 | |
| <Np> (cm$^{-3}$) | 10.7 | 5.2 | 5.9 | Over the MC interval |
| Max Np (cm$^{-3}$) | 29.2 | 12.3 | 18.4 | |
| Bt (nT) | 19.1 | 14.6 | 11.1 | Peak values |
| MC Vc (km/s) | 406 | 358 | 358 | Central speed |
| –Bz (nT) | 15.8 | 13.9 | 9 | Peak values GSE |
| –VBz (km/s nT) | 6617 | 4962 | 3147 | Peak values |
| –Dst (nT) | 175 | 80 | 87 | Peak values |

### 3.4 Ring current energy from simulations

In order to test the above conclusion that the density increase inside the MC while Bz <0 is responsible for the stronger geomagnetic storm, we perform a numerical simulation experiment to compute the total ring current energy (RCE) contrasting the low- and high-density situations. First we obtain the RCE for the three storms because they represent different densities in the MCs. Second we reduce the density inside the 2018 MC to the value in the first half and then obtain the RCE. For this purpose, we make use of the Comprehensive Inner Magnetosphere-Ionosphere (CIMI) model (Fok et al., 2014). CIMI is a kinetic model that computes the energetic ion (0.1 keV - 500 keV) and electron (1 keV - 5 MeV) distributions, plasmaspheric densities, Region 2 field-aligned currents, and subauroral ionospheric potentials. The model is a further development of the Comprehensive Ring Current Model (CRCM; Fok et al., 2001) with the addition of Radiation Belt Environment (RBE) model (Fok et al., 2011). As for CRCM, the CIMI model solves three major equations: bounce-averaged Boltzmann equation for the distribution functions of energetic ions and electrons; conservation equation of plasmasphere particles; and the ionospheric current conservation equation for the ionospheric potential. Wave-particle interactions, losses due to charge exchange and loss cone are considered. The CIMI model can be run in empirical models of magnetic field, e.g., T04 model (Tsyganenko and Sitnov, 2005) and plasma sheet models (Ebihara and Ejiri, 2000; Borovsky et al., 1998; Tsyganenko and Mukai, 2003). The CIMI model can also be coupled with MHD models, such as the BATSRUS model (e.g., Glocer et al., 2013).

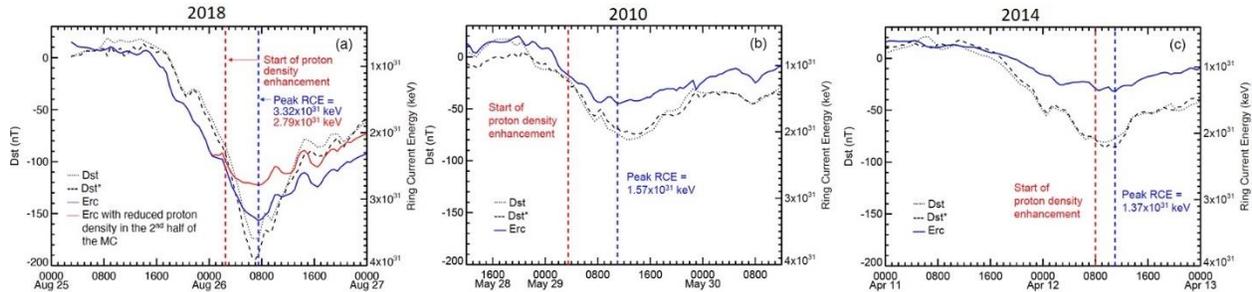

**Figure 9.** Plots of the Dst and its pressure-corrected version Dst* along with the CIMI-calculated ring current energy (Erc) for the three storms: (a) 2018 August 26, (b) 2010 May 29, and (c) 2014 April 11 (blue curves). On the right-side Y-axis, RCE increases downwards. The red and blue vertical lines mark the start of the solar wind proton density enhancement and the time of peak RCE, respectively. In (a), the red curve represents the RCE when there is no density enhancement in the second half of the MC. The lower proton density results in a lower RCE. Using the solar wind parameters shown in Figs. 5, 7, and 8 in the CIMI model, we compute RCE for the three events. Fig. 9 shows the time evolution of RCE along with Dst and its pressure-corrected version, Dst*. We see that RCE peaks at a much higher value ($3.32\times10^{31}$ keV) for the 2018 storm than that in the 2010 ($1.57\times10^{31}$ keV) and 2014 ($1.37\times10^{31}$ keV) storms. On the other hand, the peak RCE is similar in the latter two events. The steepening of the Dst profile in the 2018 storm (see Fig. 5) coincident with the density enhancement is also reflected in the RCE profile. Even the minor density enhancement in the beginning of the 2010 storm has a corresponding steepening in Dst and RCE. Even though Bz magnitude is relatively small (–9 nT) in the 2014 storm, the density enhancement towards the end of the MC increases the storm strength on par with that of the 2010 storm. The CIMI simulation thus confirms that the density enhancement is the main cause of the increased strength of the 2018 storm.

In the next CIMI run, we artificially replace the density in the back of the 2018 MC by that in the first half of the MC keeping all other solar wind parameters the same. The result is shown by the red curve in Fig. 9a. In the first half of the MC, the blue and red curves are identical. The red curve shows that the RCE ($2.79\times10^{31}$ keV) is lower by ~16% than the RCE in the actual density case ($3.32\times10^{31}$ keV). This result further confirms the importance of density inside MCs.

## 4 Discussion

The purpose of this study is to investigate the unusual circumstances that led to the third largest geomagnetic storm of solar cycle that occurred on 2018 August 26. We consider three key factors. First, the solar eruption from a quiescent filament region is extremely weak. The eruption signature is discerned from a faint PEA that persisted for more than a day. The associated white-light CME is very slow, continuing to accelerate for a day and finally becoming a typical MC. Second, the MC arrived as a unipolar cloud (FS) with its axis pointing southward, in contrast to the near-Sun indicators such as the tilt of PIL, core dimming regions, and the GCS flux rope fitted to coronagraph images. This indicates that the flux rope axis undergoes a large and complex rotation during its coronal and interplanetary propagation and the resulting configuration is conducive for reconnection with Earth's magnetic field. Third, the empirical relations that based on the high correlation between Dst and VBz fail to predict the strength of the storm. A new empirical relation between the observed Dst and the time integral of the ring current term that includes the solar wind dynamic pressure is obtained, with which the storm in hand agrees quite well. In the following we discuss some additional points related to these three considerations.

## 4.1 Evolution of the flux rope size

The kinematic analysis in combination with the reconnected flux and the PEA intensity shown in Fig. 4 suggests that the flux rope is not fully formed until it reaches a heliocentric distance of ~50 Rs. Therefore, the flux rope size obtained from the coronagraph images is not expected to be the final size. Furthermore, the assumption of self-similar expansion is also not expected to be valid in this distance range. A cylindrical flux rope fit to in-situ data using the Lepping et al. (1990) gives a flux rope radius (R) at 1 au as 0.13 au, which indicates an aspect ratio $\kappa$ = R/(Rtip – R) = 0.15. While such a $\kappa$ value (0.19) is indicated by the GCS fit to LASCO/C2 and STA/COR2 data, it increases to 0.35 in the HI-1 FOV at ~64 Rs. It is possible that the flux rope compacted after the dipolarization of the last reconnected field lines (Welsch 2018), which might have happened when the flux rope is at the outer edge of the HI-1 FOV. Assuming that the flux rope stabilizes by Rtip = 75 Rs, we can estimate the flux rope radius at this distance from the 1-au value assuming self-similar expansion. For $\kappa$ = 0.15, R =9.8 Rs at Rtip = 75 Rs and from the axial field strength $B_0$ = 23.8 nT of the flux rope fitted to in-situ data, we estimate $B_0$ at 75 Rs as 193.8 nT or 1.9 mG. This is consistent with the average $B_0$ = 52 mG at Rtip = 10 Rs (Gopalswamy et al. 2015b). From the fitted flux-rope R and $B_0$ at 1 au, we can estimate the poloidal flux as $5.8 \times 10^{21}$ Mx, which is a factor of a few larger than the observed total RC flux ($1.6 \times 10^{21}$ Mx). The correlation between 1-au poloidal flux and the RC flux has a large scatter, so the agreement is not too bad. For example, the RC flux ($1.5 \times 10^{21}$) of the 1999 April 13 CME is smaller than the poloidal flux ($5.35 \times 10^{21}$ Mx) of the associated MC (1999 April 16) observed at 1 au (Gopalswamy et al. 2018). This analysis shows that in slowly accelerating CMEs, $\kappa$ changes its value while the reconnection is ongoing and the self-similar expansion becomes valid only after the reconnection ends.

## 4.2 The effect of the nearby coronal holes on CME rotation

Weak eruptions from quiescent filament regions have been discussed before. A notable example is the eruptions on 1997 January 10–11 (Burlaga et al. 1998; Webb et al. 1998). The associated magnetic cloud results in only a moderate storm with Dst = –78 nT. The present event is even weaker at the Sun yet produced an intense geomagnetic storm that is more than two times stronger. Unlike the 1997 January event, our event has a high inclination MC, which ensures Bz<0 for an extended period of time. The high inclination compared to the tilt near the Sun indicates a large rotation of the MC between the Sun and Earth (see Chen et al. 2019 for details). Magnetic flux ropes can rotate due to internal (Fan and Gibson, 2004; Török et al, 2004; Lynch et al., 2009) and external forces (Nieves-Chinchilla et al. 2012; Kay et al. 2017). The complex rotation in our event can be attributed to the two coronal holes CH-E and CH-W shown in Fig. 1 that seem to deflect the CME in opposite directions early on at the northern and southern ends. Deflection by coronal hole magnetic fields has been documented extensively (Gopalswamy et al. 2009b and references therein). The distribution of the coronal holes at different distances and magnetic field strengths indicates external differential magnetic forces along the CME axis leading to a torque about the CME nose. In the interplanetary medium, the fast winds from the two coronal holes might have interacted with the CME causing further rotation of the flux rope.

### 4.3 The effect of the density enhancement

Farrugia et al. (1998) compare three MCs with similar solar wind profiles, including enhanced densities in the second half of the clouds. These are the MCs on 1995 October 18, 1996 May 27, and 1997 January 10 with maximum densities of 60 $cm^{-3}$, 30 $cm^{-3}$, and 185 $cm^{-3}$, respectively. Unlike our event, these are south-north MCs, so the Bz <0 part of the MCs is in the front of the MCs, with no overlap with the density enhancement. The Bz <0 part resulted in weak to intense geomagnetic storms: Dst = –127 nT (1995 October 18), –33 nT (1996 May 27), and –64 nT (1997 January 10). Therefore, the enhanced MC density does not affect the ring current (Farrugia et al. 1998; Jordanova et al. 1998) and the storm strength is simply ordered by the interplanetary electric field, VBz. The VBz in our event (~ –6500 km/s nT) is similar to that in the 1997 January 10 MC (6900 km/s nT), but our storm is almost three times more intense (–175 nT vs. –64 nT for the 1997 January 10 event). The primary difference is that the high density in the MC occurred during the Bz<0 portion of the MC. Unlike the above three events, our MC is of FS type, so Bz<0 condition prevails throughout the MC including the high-density interval and hence the enhancement of the ring current energy. Bisoi et al. (2016) report on a fully southward (FS) MC that occurred on 1998 May 2. The MC has a density enhancement in the back of the MC with several pulses. The SYM-H remains > –60 nT during these pulses. The SYM-H index also shows pulses corresponding to the density pulses, indicating that the density enhancement plays a role in the geoeffectiveness of MC substructures. After each density pulse the storm temporarily strengthens for ~1 hr.

Fenrich and Luhmann (1998) report about 40–45% the 27 MCs have of trailing density enhancement, which they identify due to compression by the following high-speed stream. They find an increased geoeffectiveness of north-south (N-S) polarity clouds due to both an increased solar wind dynamic pressure and a compressed southward field due to a high-speed solar wind stream that follows the MC. The three MCs in our study are of FS type, so the Bz <0 condition is satisfied as in the N-S MCs of Fenrich and Luhmann (1998). Following the work by Murayama (1982), Fenrich and Luhmann (1998) modified the ring current injection Q (nT/hr) in the Burton's equation (Burton et al. 1975) to include a factor $P_f^{1/3}$. Wang et al. (2003b) further modified Q by optimizing the exponent $\gamma$ and a threshold $P_f$ ($P_0$) as follows:

$$Q\,(t) = -4.4\,(VBs - 0.49)(P_f/P_0)^\gamma\,,\ VBs > 0.49\ mV/m, \qquad (3)$$

with $Q = 0$ for $VBs \geq 0.49$ mV/m. Here, Bs is the southward component defined as: $Bs = -Bz$ when $Bz < 0$ and $Bs = 0$ when $Bz \geq 0$. Wang et al. (2003b) suggest $\gamma = 0.2$ and $P_0 = 3$ nPa as optimal values to be used in Equation 3 and find that Q is the important term in the main phase of a storm. Using $\gamma = 0.5$ in Equation. 3 Xie et al. (2008) demonstrate that the Dst peak value is higher by up to 26% when there is an enhancement of $P_f$ during the main phase of a storm. Le et al. (2020) also used $\gamma = 0.5$ to find that the time integral of Q over the main phase of a storm (I (Q)) is highly correlated with the storm strength measured by the minimum value of the SYM-H index (SYM-$H_{min}$). Zhao et al. (2021) find even a better correlation between I (Q) with $\gamma = 0.5$ and $\Delta$SYM-H, the change in SYM-H over the main phase: for a set of 17 very intense storms ($\Delta$SYM-H $\leq -200$ nT) they find a correlation coefficient r =0.94. If we use the observed minimum Dst instead of SYM-$H_{min}$ the correlation remains the same for the 17 events. Xie et al. (2008), Le et al. (2020), and Zhao et al. (2021) allow a higher weightage ($\gamma = 0.5$) for the dynamic pressure in Q than the one ($\gamma = 0.2$) suggested by Wang et al. (2003b). Here we compare the effect of using $\gamma = 0.5$ vs. $\gamma = 0.2$, denoting the corresponding integrals as I (Q05) and I (Q02). We use all the 32 events listed in Zhao et al. (2021) selected by the criterion $\Delta$SYM-H $\leq -100$ nT. The 32 events are listed in Table 2 (date and Dst are as in Zhao et al.). The first 17 events are very intense ($\Delta$SYM-H $\leq -200$ nT). Also listed in the table are I (Ey), I (Q02), I (Q05), and the location of the Bz <0 interval (sheath, cloud or CIR). Figure 10 shows the scatter plot between I(Q) and Dst for the sets of 32 and 17 events with $\gamma = 0.5$ and $\gamma = 0.2$. The correlations are slightly better when $\gamma = 0.5$ for both the data sets. Higher $\gamma$ increases the weight of the dynamic pressure in Q in Equation 3. Furthermore, the correlations are almost the same for the 17 and 32 events. The high correlation indicates that most of the contribution to Dst during the main phase is due to the ring current injection, consistent with the CIMI simulation results.

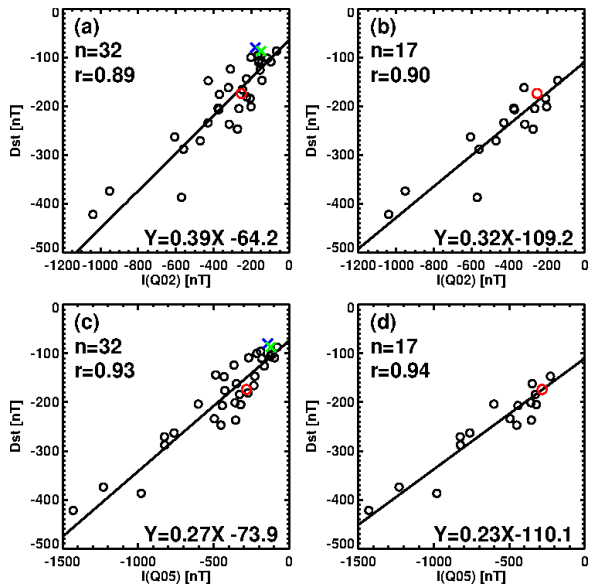

**Figure 10.** Scatter plots between Dst and I (Q) for the 32 events (left column) and 17 very intense events (right column) with $\gamma = 0.2$ (upper panel) and 0.5 (lower panel). I(Q02) and I(Q05) represent I (Q) computed with $\gamma = 0.2$ and 0.5, respectively. The Pearson's critical coefficient $r_c$ (0.297 for 32 events; 0.412 for 17 events; p =0.05) is much smaller than all the correlation coefficients (r). The red open circle represents the 2018 August 26 storm. The blue and green crosses denote the 2010 May 29 and 2014 April 11 storms. The red data point is included in the correlation, while the crosses are not.

The correlation between I (Ey) and Dst is also significant. A scatter plot between I (Ey) and Dst (not shown) yields a relation: Dst = −0.45 I(Ey) − 81.57 with r =0.80 for 32 events. The correlation is slightly better when 17 events are used (r = 0.83). The I (Ey) - Dst correlation is much weaker than the I(Q) - Dst correlation (r = 0.93), further indicating the importance of the solar wind density via Q.

**Table 2.** List of 32 storms considered for correlation analysis, the first 17 being very intense

| No. | Storm Date | Dst nT | I (Ey) Wb/m | −I (Q02) nT | −I (Q05) nT | Bz<0 Location |
|---|---|---|---|---|---|---|
| 1 | 1998/05/04 | −205 | 197 | 266 | 321 | sheath |
| 2 | 1998/09/25 | −207 | 292 | 374 | 443 | sheath |
| 3 | 1999/10/22 | −237 | 253 | 318 | 354 | cloud |
| 4 | 2000/04/06 | −288 | 370 | 560 | 824 | sheath |
| 5 | 2000/08/12 | −234 | 338 | 430 | 495 | cloud |
| 6 | 2000/09/17 | −201 | 118 | 202 | 358 | sheath |
| 7 | 2001/03/31 | −387 | 340 | 571 | 980 | cloud |
| 8 | 2001/04/11 | −271 | 277 | 471 | 826 | sheath |
| 9 | 2001/10/21 | −184 | 132 | 207 | 328 | sheath |
| 10 | 2003/11/20 | −422 | 717 | 1040 | 1431 | cloud |
| 11 | 2004/11/08 | −374 | 679 | 952 | 1231 | cloud |
| 12 | 2004/11/09 | −263 | 446 | 606 | 762 | cloud |
| 13 | 2005/05/15 | −247 | 164 | 274 | 452 | sheath |
| 14 | 2006/12/15 | −162 | 271 | 322 | 349 | cloud |
| 15 | 2015/06/22 | −204 | 247 | 377 | 602 | sheath |
| 16 | 2018/08/25 | −175 | 214 | 253 | 283 | cloud |
| 17 | 2000/05/24 | −147 | 96 | 144 | 228 | sheath |
| 18 | 2003/05/29 | −144 | 120 | 226 | 487 | sheath |
| 19 | 2003/08/17 | −148 | 390 | 429 | 431 | cloud |
| 20 | 2002/11/20 | −87 | 56 | 67 | 79 | cloud |
| 21 | 2002/10/01 | −176 | 304 | 370 | 426 | cloud |
| 22 | 2002/09/07 | −181 | 176 | 226 | 273 | sheath |
| 23 | 2002/09/04 | −109 | 87 | 97 | 97 | CIR |
| 24 | 2002/08/21 | −106 | 179 | 161 | 124 | cloud |
| 25 | 2002/08/02 | −102 | 106 | 115 | 113 | sheath |
| 26 | 2002/05/23 | −109 | 84 | 144 | 268 | sheath |
| 27 | 2002/05/11 | −110 | 140 | 162 | 182 | cloud |
| 28 | 2002/04/18 | −124 | 268 | 311 | 366 | cloud |
| 29 | 2002/03/24 | −100 | 191 | 203 | 214 | sheath |
| 30 | 2000/01/23 | −96 | 140 | 166 | 188 | cloud |
| 31 | 2001/10/03 | −166 | 228 | 248 | 234 | cloud |
| 32 | 2000/10/29 | −126 | 133 | 154 | 165 | cloud |

Figure 11 shows the time evolution of $P_f$, Ey, and Q along with the time integrals Ey and Q. There are two Q curves one with $\gamma = 0.2$ (orange) and the other with $\gamma = 0.5$ (red). There is clear sharp increase in |Q| when there is an increase in $P_f$. The peak values of |Q| in all three events coincide with peaks in $P_f$. We also see that $|Q05| > |Q02|$ whenever $P_f > P_0$ (3 nPa). The I(Q05) values for the 2018, 2010, and 2014 storms are: −283 nT, −142 nT, and −121 nT, respectively. The latter two I(Q05) are similar and much smaller than I(Q05) of the 2018 event, similar to the ordering in the total RCE and in the Dst index (see Fig. 9). The I (Q02) values follow the same pattern among the three events. On the other hand, I (Ey) is not very different among the three events: 214 Wb/m, 196 Wb/m, and 173 Wb/m for the above three events. For example, I (Ey) in the 2018 storm is higher than that in the 2010 storm only by 9%, whereas the storm strength doubles. This further demonstrates the importance of the dynamic pressure in Q. The I(Q) values of the three events in Figs. 5, 7, and 9 and the corresponding Dst values are plotted in Fig. 10. We see that the events agree with the regression line.

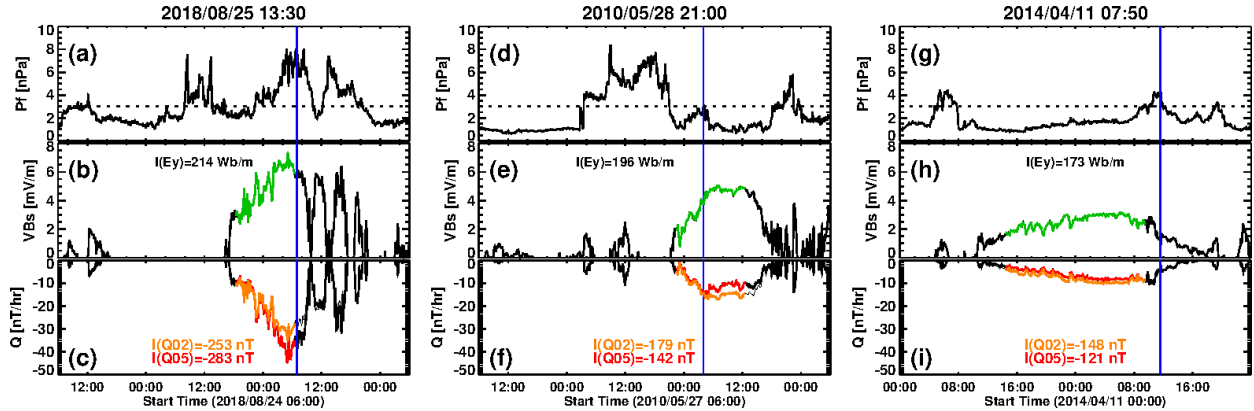

**Figure 11.** Plots of the dynamic pressure $P_f$ (a, d, g), solar wind electric field (GSM) Ey = VBs (b, e, h), and Q (c, f, i) for the 2018 August 25, 2010 May 28 and 2014 April 11 MCs. The green curves represent Ey = VBs. The orange and red curves denote and Q values with $\gamma$ = 0.2 and 0.5, respectively. The time-integrated quantities I (Ey), I (Q02), and I (Q05) are noted on the plots. The vertical blue lines mark the $P_f$ peaks for reference. Note that the peaks in Q lie within the intervals of $P_f$ (density) enhancement.

An important point to note in Table 1 is that the number of storms caused by shock sheaths and MCs (or the driving magnetic ejecta) are roughly equal: 17 cloud storms compared to 14 sheath storms. One intense storm is caused by a CIR. This indicates that the storm main phase is primarily determined by the solar wind parameters irrespective of the nature of the interplanetary structure that impacts Earth. The density/dynamic pressure variability is generally more dramatic in shock sheaths.

### 4.4 Origin of the dense material

High densities in ICMEs occur in two places: the compressed sheath ahead of the CME flux rope and inside the flux rope. The sheath comprises of heliospheric plasma and magnetic field compressed by the shock (Gosling and McComas 1987; Tsurutani et al. 1988; Kilpua et al. 2017; Meng et al. 2019). Typically, the sheath density is higher than the cloud density by a factor of ~2 (Gopalswamy et al. 2015b, their Tables 1 and 2). The Bz component is often fluctuating in the sheath interval (Tsurutani et al. 1988; Kilpua et al. 2013) and has the potential to cause time structure in Dst. The high-density material inside ICMEs can be due to compression by a high-speed stream that follows the ICME (Fenrich and Luhmann 1998) or due to eruptive prominence core of many CMEs (Fisher et al. 1981; Illing and Hundhausen 1986; Gopalswamy et al. 2003) that propagate to 1au. In-situ observations show prominence material inside ICMEs (Burlaga et al. 1998; Gopalswamy et al. 1998; Reinard 2008; Lepri and Zurbuchen 2010; Gilbert et al. 2012; Gruesbeck et al. 2012; Sharma and Srivastava 2012; Sharma et al. 2013; Gopalswamy 2015;

Mishra and Srivastava, 2015; Wang et al. 2018). The intervals of high-density prominence material are the coolest within MCs and show low Fe and O charge states. Wang et al. (2018) find that at least 27 of the 76 MCs (or 36%) they examined contain prominence material indicated by the unusual $O^{5+}$ and/or $Fe^{6+}$ abundances and in the majority of cases the prominence material is at the back end of MCs. However, occasionally azimuthal flows can redistribute the prominence material within CMEs (Kozyra et al. 2013; Manchester et al. 2014). A recent study finds that among a set of 95 isolated geomagnetic storms caused by ICMEs, the MC type ICMEs with prominence material are the most geoeffective (Li and Yao 2020). In the 2018 August MC, data on low charge states are not available, so we cannot confirm the filament material, although circumstantial evidence points to the filament material (high-density material in the coldest part of the MC). Figure 12 presents the available charge state data from the Advanced Composition Explorer (ACE) and Wind. The $O^{7+}/O^{6+}$ ratio in the slow solar wind is ~0.3 (Zhao et al. 2009). In the MC interval, the ratio increases above the slow solar wind value peaking at ~0.5 in the high-density interval. Figure 12b shows that the average Fe charge state (<QFe>) is in the range 10 to 10.9 within the MC interval with a slightly lower range in the high-density interval. These values are below the typical slow solar wind value of 11 (Lepri et al. 2001). Thus, Fe and O charge state signatures are not significant in the MC interval. This may be due to the extremely weak eruption that may not have injected hot plasma into the MC. In some filament eruption events, <QFe> and $O^{7+}/O^{6+}$ dip below the corresponding slow solar wind values at intervals corresponding to the filament material, while signatures of low Fe charge state material are pronounced in the interval (Gopalswamy 2015b). The 2018 August 25 MC does not show this dip. We cannot say whether low Fe charge states are enhanced in the high-density interval because such data are not available any longer. The only hint of prominence material comes from the $He^{++}$ signature, which shows a sharp increase (from 0.4 to 1 $cm^{-3}$) within the high-density interval lasting for about an hour. Additionally, there is the possibility that the large filament fragment present at the northern end of the channel (see Fig. 2) and erupting within the acceleration phase of the CME would have found its way to the back of the flux rope. The high-speed stream that follows our MC can also compress the material at the back of the MC. Irrespective of the origin of the high-density material, its influence on the geoeffectiveness is significant. Further progress in understanding the high-density material in ICMEs can be made by considering MCs with high-density material but not followed by a high-speed stream.

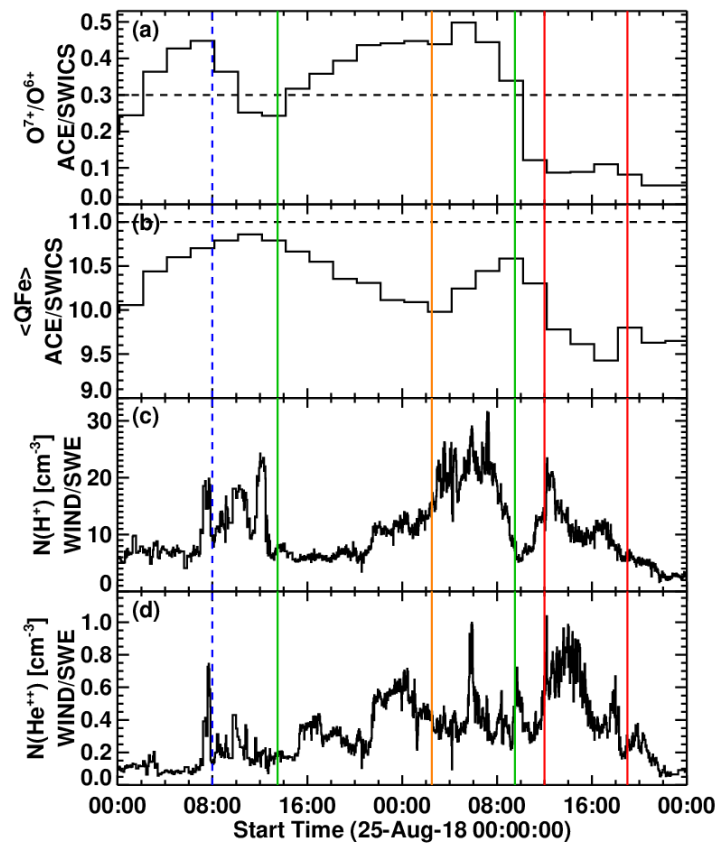

Figure 12. Charge state data from ACE/SWICS and Wind/SWE during the 2018 August 25 MC. (a) the ratio ($O^{7+}/O^{6+}$) of the number densities of $O^{7+}$ and $O^{6+}$ ions, (b) the average Fe charge state (<QFe>), (c) proton density, and (d) the density of $He^{++}$. The blue vertical dashed line marks the rear boundary of MC1. The MC in question is between the green vertical lines as in Fig. 5. The CIR interval is between the red vertical lines. The proton density enhancement is between the orange line and the second green line.

## 5 Conclusions

We investigated the solar and interplanetary causes behind the third largest geomagnetic storm of solar cycle 24 that occurred on 2018 August 26. The high intensity of the storm is result of the combined occurrence of prolonged CME acceleration, complex CME rotation, and the presence of high-density material in the back of the MC. The solar source is a quiescent filament channel containing filament fragments. The eruption of the filament channel is accompanied by a slow CME, twin case dimming, and a post eruption arcade, typical of most eruptions. The CME acceleration lasted for a day until the CME reached a heliocentric distance of ~50 Rs. The continued acceleration is powered by magnetic reconnection beneath the filament channel as evidenced by the correspondence among the time profiles of the CME acceleration, time derivative of the PEA intensity in EUV, and the rate of change of the reconnected flux. This is direct evidence that the CME propelling force can act at distances >50 Rs. The speed at this distance and the total reconnected flux in the eruption agree with the reconnected flux - CME speed relation. Therefore, in every respect (photospheric, chromospheric, and coronal) the CME behaves like a normal CME, so it is probably not a good idea to designate it as a stealth CME. The one exception is the complete absence of nonthermal radio signatures. The prolonged acceleration results in a 1-au speed exceeding the slow solar wind.

Comparison among the tilt angles of the photospheric neutral line, filament channel, the lines connecting the dimming regions, axis of the GCS flux rope, and axis of the 1-au MC point to a complex rotation of the CME flux rope between the Sun and Earth. We suggest that the multiple coronal holes located near the filament channel creates a situation where differential magnetic forces act on the flux rope axis causing deflections of different extent at different locations. The net result is an early counterclockwise rotation. In the interplanetary medium, the solar wind from the coronal hole on the east side of the filament channel is likely to have pushed the northern part of the CME westward, resulting in the clockwise rotation and hence the high inclination of the MC.

We find a significant steepening of the Dst time profile coincident with the increase in density inside the MC interval. The steepening results in a significantly stronger storm strength as compared to cases without the high-density material inside MCs. The 2018 storm is also significantly larger than what is predicted by empirical formulas of Dst (up to a factor of 2) that do not take into account of the solar wind density. Complex time profiles of the Dst index in the storm main phase can occur when the dynamic pressure and Ey vary. Under Bz < 0 condition, the dynamic pressure primarily defines the time profile. When the dynamic pressure is low, Ey defines the time structure. Both of these are affected by the solar wind density. The total ring current energy obtained from the CIMI model and the time integral of the ring current injection are consistent with the high storm intensity when the solar wind dynamic pressure is incorporated into the definition of the ring current injection, Q. A comparison of the 2018 storm simulation with that of the 2010 and 2014 storms point to the enhanced proton density (and hence the dynamic pressure) inside the 2018 MC as the primary factor behind the unusually high storm intensity. We also find a high correlation (r = 0.93) between the Dst index and the integral of the ring current injection rate over the main phase for a set of 32 intense storms that occurred in solar cycles 23 and 24. The 32 storms occur during Bz <0 intervals in different types of interplanetary structures: MCs, shock sheaths, and CIR.

## Acknowledgments

This work benefited greatly from NASA's open data policy in using SDO (https://sdo.gsfc.nasa.gov/), SOHO (https://soho.nascom.nasa.gov/), STEREO (https://stereo-ssc.nascom.nasa.gov/), OMNI (https://omniweb.gsfc.nasa.gov/), and Wind (https://wind.nasa.gov/) data. STEREO is a mission in NASA's Solar Terrestrial Probes program. The Dst index used in this paper was provided by the WDC for Geomagnetism, Kyoto (http://wdc.kugi.kyoto-u.ac.jp/wdc/Sec3.html). SOHO is a project of international collaboration between ESA and NASA. Preliminary results of this investigation were presented at the International Space Weather Action Teams (ISWAT) inaugural meeting held during 10-14 February 2020 in Cape Canaveral, Florida. Work supported by NASA's Living With a Star Program. PM is supported in part by the NSF grant AGS-2043131.

## References

Abunin A. A., Abunina, M. A., Belov, A. V., & Chertok, I. M. (2020). Peculiar Solar Sources and Geospace Disturbances on 20-26 August 2018. *Solar Physics*, 295(1), 7. https://doi.org/10.1007/s11207-019-1574-8

Barnes, C. W., Simpson, J. A. (1976). Evidence for interplanetary acceleration of nucleons in corotating interaction regions. *The Astrophysical Journal*, 210, L91. https://doi.org/10.1086/182311

Belcher, J. W., Davis, L. Jr. (1971). Large-amplitude Alfvén waves in the interplanetary medium, 2. *Journal of Geophysical Research:Space Physics*, 76, 3534. https://doi.org/10.1029/JA076i016p03534

Bisoi, S. K., Chakrabarty, D., Janardhan, P., Rastogi, R. G., Yoshikawa, A., Fujiki, K., Tokumaru, M., Yan Y. (2016). The prolonged southward IMF-Bz event of 2-4 May 1998: Solar, interplanetary causes and geomagnetic consequences. *Journal of Geophysical Research: Space Physics*, 121(5), 3883–3904. https://doi.org/10.1002/2015JA022185

Borovsky, J. E., Thomsen, M. F., & Elphic R. C. (1998). The driving of the plasma sheet by the solar wind. *Journal of Geophysical Research: Space Physics*, 103(8)，17617–17640. https://doi.org/10.1029/97JA02986

Brueckner, G. E., Howard, R. A., Koomen, M. J., Korendyke, C. M., Michels, D. J., Moses, J. D., Socker, D. G., Dere, K. P., et al. (1995). The large angle spectroscopic coronagraph (LASCO), *Solar Physics*, 162, 357–402. https://doi.org/10.1007/BF00733434

Burlaga, L., Sittler, E., Mariani, F., Schwenn, R. (1981). Magnetic loop behind an interplanetary shock: Voyager, Helios, and IMP 8 observations. *Journal of Geophysical Research: Space Physics*, 86, 6673–6684. https://doi.org/10.1029/JA086iA08p06673

Burlaga, L. F., Fitzenreiter, R., Lepping, R., Ogilvie, K., Szabo, A., Lazarus, A., Steinberg, J., Gloeckler, G., et al. (1998). A magnetic cloud containing prominence material: January 1997, *Journal of Geophysical Research: Space Physics*, 103(1), 277–286. https://doi.org/10.1029/97JA02768

Burton, R. K., McPherron, R. L., & Russell, C. T. (1975). An empirical relationship between interplanetary conditions and Dst. *Journal of Geophysical Research*, 80, 4204–4214. https://doi.org/10.1029/JA080i031p04204

Chen, C., Liu, Y. D., Wang, R., Zhao, X., Hu, H., & Zhu, B. (2019). Characteristics of a Gradual Filament Eruption and Subsequent CME Propagation in Relation to a Strong Geomagnetic Storm. *The Astrophysical Journal*, 884(1), 90. https://doi.org/10.3847/1538-4357/ab3f36

Cheng, L. -B., Le, G. -M., Zhao, & M. -X. (2020). Sun-Earth connection event of super geomagnetic storm on 2001 March 31: the importance of solar wind density. *Research in Astronomy and Astrophysics*, 20(3), 036. https://doi.org/10.1088/1674-4527/20/3/36

Cliver, E. W., Kahler, S. W., Kazachenko, M., & Shimojo, M. (2019). The Disappearing Solar Filament of 2013 September 29 and Its Large Associated Proton Event: Implications for Particle Acceleration at the Sun. *The Astrophysical Journal*, 877(1), 11. https://doi.org/10.3847/1538-4357/ab0e03

Dennis, B. D., & Zarro, D. (1993). The Neupert Effect - what can it Tell up about the Impulsive and Gradual Phases of Solar Flares. *Space Physics*, 146(1), 177–190. https://doi.org/10.1007/BF00662178

Ebihara, Y., & Ejiri M. (2000). Simulation study on fundamental properties of the storm time ring current. *Journal of Geophysical Research*, 105(15), 15843–15859. https://doi.org/10.1029/1999JA900493

Fan, Y., & Gibson, S. E. (2004). Numerical Simulations of Three-dimensional Coronal Magnetic Fields Resulting from the Emergence of Twisted Magnetic Flux Tubes. *The Astrophysical Journal*, 609(2), 1123–1133. https://doi.org/10.1086/421238

Fenrich, F. R., & Luhmann, J. G. (1998). Geomagnetic response to magnetic clouds of different polarity. *Geophysical Research Letters*, 25(15), 2999–3002. https://doi.org/10.1029/98GL51180

Fisher, R., Garcia, C. J., Seagraves, P. (1981). On the coronal transient-eruptive prominence of 1980 August 5. *The Astrophysical Journal*, 246, L161–L164. https://doi.org/10.1086/183575

Farrugia, C. J., Scudder, J. D., Freeman, M. P., Janoo, L., Quinn, J. M., Arnoldy, R. L., Torbert, R. B., Burlaga, L. F., et al. (1998). Geoeffectiveness of three Wind magnetic clouds: A comparative study. *Journal of Geophysical Research*, 103(8), 17261–17278. https://doi.org/10.1029/98JA00886

Fok, M. -C., Wolf, R. A., Spiro, R. W., & Moore T. E. (2001). Comprehensive computational model of the Earth's ring current. *Journal of Geophysical Research*, 106(5), 8417–8424. https://doi.org/10.1029/2000JA000235

Fok, M. -C., Glocer, A., Zheng, Q., Horne, R. B., Meredith, N. P., Albert, J. M., & Nagai, T. (2011). Recent developments in the radiation belt environment model. *Journal of Atmospheric and Solar-Terrestrial Physics*, 73(11–12), 1435–1443. https://doi.org/10.1016/j.jastp.2010.09.033

Fok, M. -C., Buzulukova, N. Y., Chen, S. -H., Glocer, A., Nagai, T., Valek, P., & Perez, J. D. (2014). The Comprehensive Inner Magnetosphere-Ionosphere Model. *Journal of Geophysical Research*, 119(9), 7522–7540. https://doi.org/10.1002/2014JA020239

Gilbert, J. A., Lepri, S. T., Landi, E., & Zurbuchen, T. H. (2012). First measurements of the complete heavy-ion charge state distributions of C, O, and Fe associated with interplanetary coronal mass ejections. *The Astrophysical Journal*, 751(1), 20. https://doi.org/10.1088/0004-637X/751/1/20

Glocer, A., Fok, M., Meng, X., Toth, G., Buzulukova, N., Chen, S., & Lin, K. (2013). CRCM + BATS-R-US two way coupling. *Journal of Geophysical Research：Space Physics*, 118(4), 1635–1650. https://doi.org/10.1002/jgra.50221

Gonzalez, Walter D., Tsurutani, Bruce T. (1987). Criteria of interplanetary parameters causing intense magnetic storms ( Dst < -100 nT). *Planetary and Space Science*, 35, 1101. https://doi.org/10.1016/0032-0633(87)90015-8

Gonzalez, W. D., Tsurutani, B. T., Gonzalez, A. L. C., Smith, E. J., Tang, F., Akasofu, S.-I. (1989). Solar wind-magnetosphere coupling during intense magnetic storms (1978-1979). *Journal of Geophysical Research: Space Physics*, 94, 8835. https://doi.org/10.1029/JA094iA07p08835

Gonzalez, W. D., & Echer, E. (2005). Study on the peak Dst and peak negative Bz relationship during intense geomagnetic storms. *Geophysical Research Letters*, 32(18), L18103. https://doi.org/10.1029/2005GL023486

Gopalswamy, N. (2006). Coronal mass ejections and type II radio bursts. In N. Gopalswamy, R. Mewaldt, J. Torsti (Eds.), *Solar Eruptions and Energetic Particles, Geophysical Monograph Series* (Vol. 165, pp. 207–220). Washington, DC: American Geophysical Union.

Gopalswamy, N. (2012). Energetic Particle and Other Space Weather Events of Solar Cycle 24. In H. Qiang, L. Gang, G. P. Zank, X. Ao, O. Verkhoglyadova, & J. H. Adams (Eds.), *Space weather: The space radiation environment: 11the annual international astrophysics conference, AIP Conference Proceedings,* 1500(1), 14–19. https://doi.org/10.1063/1.4768738

Gopalswamy, N. (2015). The Dynamics of Eruptive Prominences, In J. -C. Vial, & O. Engvold (Eds.), *Solar Prominences, Astrophysics and Space Science Library* (Vol. 415, pp 381–409). Switzerland: Springer International Publishing. https://doi.org/10.1007/978-3-319-10416-4_15

Gopalswamy, N. (2018). Chapter 2 - Extreme Solar Eruptions and their Space Weather Consequences. In N. Buzulukova (Ed.), *Extreme Events in Geospace* (pp 37–63). Elsevier, https://doi.org/10.1016/B978-0-12-812700-1.00002-9

Gopalswamy, N., Hanaoka, Y., Kosugi, T., Lepping, R. P., Steinberg, J.T., Plunkett, S., Howard, R.A., Thompson, B. J., et al. (1998). On the relationship between coronal mass ejections and magnetic clouds. *Geophysical Research Letters*, 25(14), 2485–2488. https://doi.org/10.1029/98GL50757

Gopalswamy, N. , Shimojo, M., Lu, W. , Yashiro, S., Shibasaki, K., Howard, R. A. (2003). Prominence Eruptions and Coronal Mass Ejection: A Statistical Study Using Microwave Observations. *The Astrophysical Journal*, 586, 562. https://doi.org/10.1086/367614

Gopalswamy, N., Yashiro, S., Michalek, G., Xie, H., Lepping, R. P., & Howard, R. A. (2005). Solar source of the largest geomagnetic storm of cycle 23. *Geophysical Research Letters*, 32(12), L12S09. https://doi.org/10.1029/2004GL021639

Gopalswamy, N., Akiyama, S., Yashiro, S., Michalek, G., & Lepping R. P. (2008). Solar sources and geospace consequences of interplanetary magnetic clouds observed during solar cycle 23. *Journal of Atmospheric and Solar-Terrestrial Physics*, 70(2–4), 245–253. https://doi.org/10.1016/j.jastp.2007.08.070

Gopalswamy, N., Yashiro, S., Michalek, G., Stenborg, G., Vourlidas, A., Freeland, & S., Howard, R. (2009a). The SOHO/LASCO CME Catalog. *Earth Moon Planets*, 104(1–4), 295–313. https://doi.org/10.1007/s11038-008-9282-7

Gopalswamy, N., Mäkelä, Xie, H., Akiyama, S., Yashiro, S. (2009b). CME interactions with coronal holes and their interplanetary consequences. *Journal of Geophysical Research: Space Physics*, 114 (A3), CiteID A00A22, https://doi.org/10.1029/2008JA013686

Gopalswamy, N., Xie, H., Mäkelä, P., Akiyama, S., Yashiro, S., Kaiser, M. L., Howard, R. A., & Bougeret J. -L. (2010). Interplanetary shocks lacking type II radio bursts. *The Astrophysical Journal*, 710(2), 1111–1126. https://doi.org/10.1088/0004-637X/710/2/1111

Gopalswamy, N., Akiyama, S., Yashiro, S., Xie, H., Mäkelä, P., & Michalek, G. (2015a). *The Mild Space Weather in Solar Cycle 24*. Paper presented at 14th International Ionospheric Effects Symposium on 'Bridging the gap between applications and research involving ionospheric and space weather disciplines' May 12-14, Alexandria, VA. Retrieved from https://arxiv.org/abs/1508.01603

Gopalswamy, N., Yashiro, S., Xie, H., Akiyama, S., & Mäkelä P. (2015b). Properties and geoeffectiveness of magnetic clouds during solar cycles 23 and 24. *Journal of Geophysical Research: Space Physics*, 120(11), 9221–9245. https://doi.org/10.1002/2015JA021446

Gopalswamy, N., Yashiro, S., & Akiyama, S. (2015c). Kinematic and Energetic Properties of the 2012 March 12 Polar Coronal Mass Ejection. *The Astrophysical Journal*, 809(1), 106. https://doi.org/10.1088/0004-637X/809/1/106

Gopalswamy, N., Mäkelä, P., Akiyama, S., Yashiro, S., Xie, H., Thakur, N., & Kahler, S. W. (2015d). Large Solar Energetic Particle Events Associated with Filament Eruptions Outside of Active Regions. *The Astrophysical Journal*, 806(1), 8. https://doi.org/10.1088/0004-637X/806/1/8

Gopalswamy, N., Akiyama, S., Yashiro, S., & Xie, H. (2018). Coronal Flux Ropes and their Interplanetary Counterparts. *Journal of Atmospheric and Solar-Terrestrial Physics*, 180, 35–45. https://doi.org/10.1016/j.jastp.2017.06.004

Gosling, J. T. (1993). The solar flare myth. *Journal of Geophysical Research*, 98, 18937. https://doi.org/10.1029/93JA01896

Gosling, J. T., Hundhausen, A. J., Pizzo, V., Asbridge, J. R. (1972). Compressions and rarefactions in the solar wind: Vela 3. *Journal of Geophysical Research*, 77(28), 5442–5454. https://doi.org/10.1029/JA077i028p05442

Gosling, J. T., McComas, D. J. (1987). Field line draping about fast coronal mass ejecta: A source of strong out-of-the-ecliptic interplanetary magnetic fields. *Geophysical Research Letters*, 14, 355. https://doi.org/10.1029/GL014i004p00355

Gruesbeck, J. R., Lepri, S. T., & Zurbuchen, T. H. (2012). Two-plasma model for low charge state interplanetary coronal mass ejection observations. *The Astrophysical Journal*, 760(2), 141. https://doi.org/10.1088/0004-637X/760/2/141

Howard, R. A., Moses, J. D., Vourlidas, A., Newmark, J. S., Socker, D. G., Plunkett, S. P., Korendyke, C. M., Cook, J. W. et al. (2008). Sun Earth Connection Coronal and Heliospheric Investigation (SECCHI). *Space Science Reviews*, 136(1–4), 67–115. https://doi.org/10.1007/s11214-008-9341-4

Illing, R. M. E., Hundhausen, A. J. (1986). Disruption of a coronal streamer by an eruptive prominence and coronal mass ejection. *Journal of Geophysical Research*, 91, 10951. https://doi.org/10.1029/JA091iA10p10951

Jordanova, V. K., Farrugia, C. J., Janoo, L., Quinn, J. M., Torbert, R. B., Ogilvie, K. W., Lepping, R. P., Steinberg, J. T., et al. (1998). October 1995 magnetic cloud and accompanying storm activity: Ring current evolution. *Journal of Geophysical Research*, 103(1), 79–92. https://doi.org/10.1029/97JA02367

Jordanova, V. K., Kistler, L. M., Thomsen, M. F., & Mouikis C. G. (2003). Effects of plasma sheet variability on the fast initial ring current decay. *Geophysical Research Letters*, 30(6), 1311. https://doi.org/10.1029/2002GL016576

Kahler, S. W., Cliver, E. W., Cane, H. V., McGuire, R. E., Stone, R. G., & Sheeley, N. R., Jr. (1986). Solar Filament Eruptions and Energetic Particle Events. *The Astrophysical Journal*, 302, 504–510. https://doi.org/10.1086/164009

Kakad, B., Kakad, A., Ramesh, D. S., & Lakhina, G. S. (2019). Diminishing activity of recent solar cycles (22–24) and their impact on geospacer. *Journal of Space Weather Space Climate*, 9, A1. https://doi.org/10.1051/swsc/2018048

Kane, R. P. (2005). How good is the relationship of solar and interplanetary plasma parameters with geomagnetic storms? *Journal of Geophysical Research: Space Physics*, 110(2), A02213. https://doi.org/10.1029/2004JA010799

Kay, C., Gopalswamy, N., Xie, H., & Yashiro, S. (2017). Deflection and Rotation of CMEs from Active Region 11158. *Solar Physics*, 292(6), 78, https://doi.org/10.1007/s11207-017-1098-z

Kilpua, E. K. J., Hietala, H., Koskinen, H. E. J., Fontaine, D., Turc, L. (2013). Magnetic field and dynamic pressure ULFfluctuations in coronal-mass-ejection-driven sheath regions. *Ann Geophys,* 31, 1559. https://doi.org/10.5194/angeo-31-1559-2013

Kilpua, E., Koskinen, H., & Pulkkinen T. I. (2017). Coronal mass ejections and their sheath regions in interplanetary space. *Living Reviews in Solar Physics*, 14(1), 5. https://doi.org/10.1007/s41116-017-0009-6

Kozyra, J. U., Manchester, W. B., Escoubet, C. P., Lepri, S. T., Liemohn, M. W. Gonzalez, W. D., Thomsen, M. W., & Tsurutani, B. T. (2013). Earth's collision with a solar filament on 21 January 2005: Overview. *Journal of Geophysical Research: Space Physics*, 118(10), 5967–5978. https://doi.org/10.1002/jgra.50567

Le, G.-M., Liu, G.-X., & Zhou, M.-M. (2020). Dependence of Major Geomagnetic Storm Intensity (Dst≤−100 nT) on Associated Solar Wind Parameters. *Solar Physics*, 295, 108. https://doi.org/10.1007/s11207-020-01675-3

Lemen, J. Title, A., Akin, D., Boerner, P. F., Chou, C., Drake, J. F., Duncan, D. W., Edwards, C. G., et al. (2012). The Atmospheric Imaging Assembly (AIA) on the Solar Dynamics Observatory (SDO). *Solar Physics*, 275(1–2), 17–40. https://doi.org/10.1007/s11207-011-9776-8

Lepping, R. P., Burlaga, L. F., & Jones, J. A. (1990). Magnetic field structure of interplanetary magnetic clouds at 1 AU. *Journal of Geophysical Research*, 95(8), 11957–11965. https://doi.org/10.1029/JA095iA08p11957

Lepri, S. T., Zurbuchen, T. H., Fisk, L. A., Richardson, I. G., Cane, H. V., Gloeckler, G. (2001). Iron charge distribution as an identifier of interplanetary coronal mass ejections. *Journal of Geophysical Research*, 106 (A12), 29231. https://doi.org/10.1029/2001JA000014

Lepri, S.T., & Zurbuchen, T. H. (2010). Direct observational evidence of filament material within interplanetary coronal mass ejections. *The Astrophysical Journal Letters*, 723(1), L22–L27. https://doi.org/10.1088/2041-8205/723/1/L22

Li, D., & Yao, S. (2020). Stronger Southward Magnetic Field and Geoeffectiveness of ICMEs Containing Prominence Materials Measured from 1998 to 2011. *The Astrophysical Journal*, 891(1), 79. https://doi.org/10.3847/1538-4357/ab7197

Liu, Y. D., Hu, H., Wang, R., Yang, Z., Zhu, B., Liu, Y. A., Luhmann, J. G., & Richardson, J. D. (2015). Plasma and Magnetic Field Characteristics of Solar Coronal Mass Ejections in Relation to Geomagnetic Storm Intensity and Variability. *The Astrophysical Journal*, 809(2), L34. https://doi.org/10.1088/2041-8205/809/2/L34

Ying D. Liu, Hu, H., Wang, C., Luhmann, J. G., Richardson, J. D., Yang, Z., and Wang, R. (2016). On sun-to-earth propagation of coronal mass ejections: ii. Slow events and comparison with others. *The Astrophysical Journal Supplement Series*, 222 (2), 23. https://doi.org/10.3847/0067-0049/222/2/23

Lopez, R. E., Wiltberger, M., Hernandez, S., & Lyon, J. G. (2004). Solar wind density control of energy transfer to the magnetosphere. *Geophysical Research Letters*, 31(8), L08804. https://doi.org/10.1029/2003GL018780

Lynch, B. J., Antiochos, S. K., Li, Y., Luhmann, J. G., & DeVore, C. R. (2009). Rotation of Coronal Mass Ejections During Eruption. *The Astrophysical Journal*, 697(2), 1918–1927. https://doi.org/10.1088/0004-637X/697/2/1918

Manchester, W. B., Kozyra, J. U., Lepri, S. T., & Lavraud, B. (2014). Simulation of magnetic cloud erosion during propagation. *Journal of Geophysical Research*, 119(7), 5449–5464. https://doi.org/10.1002/2014JA019882

Martin, S. F. (1998). Conditions for the Formation and Maintenance of Filaments (Invited Review), *Solar Physics*, 182(1), 107–137. https://doi.org/10.1023/A:1005026814076

Meng, X., Tsurutani, B. T., Mannucci, A. J. (2019), The Solar and Interplanetary Causes of Superstorms (Minimum Dst ≤ −250 nT) During the Space Age. *Journal of Geophysical Research*, 124, 3926–3948, https://doi.org/10.1029/2018JA026425

Murayama, T. (1982). Coupling function between solar wind parameters and geomagnetic indices. *Reviews of Geophysics*, 20(3), 623– 629. https://doi.org/10.1029/RG020i003p00623

Mishra, W., & N. Srivastava, (2015). Heliospheric tracking of enhanced density structures of the 6 October 2010 CME. *Journal of Space Weather and Space Climate*, 5, A20. https://doi.org/10.1051/swsc/2015021

Mishra, S. K., & Srivastava, A. K. (2019). Linkage of Geoeffective Stealth CMEs Associated with the Eruption of Coronal Plasma Channel and Jet-Like Structure. *Solar Physics*, 294(12), 169. https://doi.org/10.1007/s11207-019-1560-1

Neupert, W. M. (1968). Comparison of Solar X-Ray Line Emission with Microwave Emission during Flares, *The Astrophysical Journal*, 153, L59. https://doi.org/10.1086/180220

Nieves-Chinchilla, T., Colaninno, R., Vourlidas, A., Szabo, A., Lepping, R. P., Boardsen, S. A., Anderson, B. J., & Korth, H. (2012). Remote and in situ observations of an unusual Earth-directed coronal mass ejection from multiple viewpoints, *Journal of Geophysical Research*, 117(6), A06106. https://doi.org/10.1029/2011JA017243

Nitta, N. V., Mulligan, T., Kilpua, E. K. J., Lynch, B. J., Mieria, M., O'Kane, J., Pagano, P., Palmerio, E., et al. (2021). Understanding the Origins of Problem Geomagnetic Storms Associated With "Stealth" Coronal Mass Ejections, *Space Science Review*, 217(8), 82. https://doi.org/10.1007/s11214-021-00857-0

Nose, M., Iyemori, T., Sugiura, M., & Kamei, T. (2015). World Data Center for Geomagnetism, Kyoto Geomagnetic Dst index, https://doi.org/10.17593/14515-74000

Piersanti, M., De Michelis, P., Del Moro, D., Tozzi, R., Pezzopane, M., Consolini, G., Marcucci, M. F., Laurenza M, et al. (2020). From the Sun to Earth: effects of the 25 August 2018 geomagnetic storm. *Annales Geophysicae*, 38(3), 703–724. https://doi.org/10.5194/angeo-38-703-2020

Reinard, A. A. (2008). Analysis of interplanetary coronal mass ejection parameters as a function of energetics, source location, and magnetic structure. *The Astrophysical Journal*, 682(2), 1289–1305. https://doi.org/10.1086/589322

Richardson, I. G. (2013). Geomagnetic activity during the rising phase of solar cycle 24. *Journal of Space Weather Space Climate*, 3, A08. https://doi.org/10.1051/swsc/2013031

Sachdeva, N., Subramanian, P., Colaninno, R., & Vourlidas, A. (2015). CME Propagation: Where does Aerodynamic Drag 'Take Over'? *The Astrophysical Journal*, 809(2), 158. https://doi.org/10.1088/0004-637X/809/2/158

Scherrer, P., Schou, J., Bush, R. I., Kosovichv, A. G., Bogart, R. S., Hoeksema, J. T., Liu, Y., Duvall, T. L., et al. (2012). The Helioseismic and Magnetic Imager (HMI) Investigation for the Solar Dynamics Observatory (SDO). *Solar Physics*, 275(2), 207–227. https://doi.org/10.1007/s11207-011-9834-2

Sharma, R., & N. Srivastava. (2012). Presence of solar filament plasma detected in interplanetary coronal mass ejections by in situ spacecraft. *Journal of Space Weather and Space Climate*, 2, A10. https://doi.org/10.1051/swsc/2012010

Sharma, R., Srivastava, N., Chakrabarty, D., Möstl, C., & Hu, Q. (2013). Interplanetary and geomagnetic consequences of 5 January 2005 CMEs associated with eruptive filaments. *Journal of Geophysical Research: Space Physics*, 118(7), 3954–3967. https://doi.org/10.1002/jgra.50362

Sheeley, N. R. Jr., Martin, S. F., Panasenco, O., & Warren, H. P. (2013). Using Coronal Cells to Infer the Magnetic Field Structure and Chirality of Filament Channels, *The Astrophysical Journal*, 772(2), 88. https://doi.org/10.1088/0004-637X/772/2/88

Smith, E. J. & Wolfe, J. H. (1976). Observations of interaction regions and corotating shocks between one and five AU: Pioneers 10 and 11. *Geophysical Research Letters*, 3(63), 137. https://doi.org/10.1029/GL003i003p00137

Temmer, M., Rollett, T., Möstl, C., Veronig, A. M., Vršnak, B., & Odstrčil, D. (2011). Influence of the Ambient Solar Wind Flow on the Propagation Behavior of Interplanetary Coronal Mass Ejections. *The Astrophysical Journal*, 743(2), 101. https://doi.org/10.1088/0004-637X/743/2/101

Thampi, S. V., Krishnaprasad, C., Nampoothiri, G. N., & Pant, T. K. (2021). The impact of a stealth CME on the Martian topside ionosphere, *Monthly Notices of the Royal Astronomical Society*, 503(1), 625–632. https://doi.org/10.1093/mnras/stab494

Thernisien, A. (2011). Implementation of the Graduated Cylindrical Shell Model for the Three-dimensional Reconstruction of Coronal Mass Ejections. *The Astrophysical Journal Supplement*, 194(2), 33. https://doi.org/10.1088/0067-0049/194/2/33

Török, T., Kliem, B., & Titov, V. S. (2004). Ideal kink instability of a magnetic loop equilibrium. *Astronomy and Astrophysics*, 413, L27–L30. https://doi.org/10.1051/0004-6361:20031691

Tsurutani, B. T., Smith, E. J., Gonzalez, W. D., Tang, F., Akasofu, S. I. (1988). Origin of interplanetary southward magnetic fields responsible for major magnetic storms near solar maximum (1978–1979). *Journal of Geophysical Research: Space Physics*, 93, 8519. https://doi.org/10.1029/JA093iA08p08519

Tsyganenko, N. A., & Mukai, T. (2003). Tail plasma sheet models derived from Geotail particle data. *Journal of Geophysical Research: Space Physics*, 108(A3), 1136. https://doi.org/10.1029/2002JA009707

Tsyganenko, N. A., & Sitnov M. I. (2005). Modeling the dynamics of the inner magnetosphere during strong geomagnetic storms. *Journal of Geophysical Research: Space Physics*, 110(A3), A03208. https://doi.org/10.1029/2004JA010798

Wang, Y., Shen, C. L., Wang, S., Ye, & P. Z. (2003)a. An empirical formula relating the geomagnetic storm's intensity to the interplanetary parameters: −VBz and δt. *Geophysical Research Letters*, 30(20), 2039. https://doi.org/10.1029/2003GL017901

Wang, C. B., Chao, J. K., & Lin, C. H. (2003b). Influence of the solar wind dynamic pressure on the decay and injection of the ring current. *Journal of Geophysical Research*, 108(A9), 1341. https://doi.org/10.1029/2003JA009851

Wang J., Feng, H., & Zhao, G. (2018). Cold prominence materials detected within magnetic clouds during 1998–2007. *Astronomy & Astrophysics*, 616, A41. https://doi.org/10.1051/0004-6361/201732807

Webb, D., & Nitta, N. (2017). Understanding Problem Forecasts of ISEST Campaign Flare-CME Events. *Solar Physics*, 292(10), 142. https://doi.org/10.1007/s11207-017-1166-4

Webb, D., Cliver, E., Gopalswamy, N., Hudson, H., & St. Cyr, O. C. (1998). The solar origin of the January 1997 coronal mass ejection, magnetic cloud and geomagnetic storm. *Geophysical Research Letters*, 25(14), 2469–2472. https://doi.org/10.1029/98GL00493

Weigel, R. S. (2010). Solar wind density influence on geomagnetic storm intensity. *Journal of Geophysical Research*, 115(A9), A09201. https://doi.org/10.1029/2007GL032298

Welsch, B. T. (2018). Flux Accretion and Coronal Mass Ejection Dynamics. *Solar Physics*, 293(7), 113. https://doi.org/10.1007/s11207-018-1329-y

Wilcox, J. M. & Ness, N. F. (1965). Quasi-Stationary Corotating Structure in the Interplanetary Medium. Journal of Geophysical Research: Space Physics, 170(23), 5793. https://doi.org/10.1029/JZ070i023p05793

Wu, C. -C., & Lepping, R. P. (2002). Effect of solar wind velocity on magnetic cloud-associated magnetic storm intensity. *Journal of Geophysical Research: Space Physics*, 107(A11), 1346. https://doi.org/10.1029/2002JA009396

Wu, C. -C., Liou, K., Lepping, R. P., Hutting, L., Plunkett, S., Howard, R. A., & Socker, D. (2016). The first super geomagnetic storm of solar cycle 24: The St. Patrick's day event (17 March 2015). *Earth, Planets and Space*, 68(1), 151. https://doi.org/10.1186/s40623-016-0525-y

Xie, H., Gopalswamy, N., Cyr, O. C. S., & Yashiro, S. (2008). Effects of solar wind dynamic pressure and preconditioning on large geomagnetic storms. *Geophysical Research Letters*, 35(6), L06S08. https://doi.org/10.1029/2007GL032298

Yashiro, S., Gopalswamy, N., Michalek, G., St. Cyr, O. C., Plunkett, S. P., Rich, N. B., & Howard, R. (2004). A catalog of white light coronal mass ejections observed by the SOHO spacecraft. *Journal of Geophysical Research: Space Physics*, 109(A7), A07105. https://doi.org/10.1029/2003JA010282

Yue, C., and Q. Zong (2011). Solar wind parameters and geomagnetic indices for four different interplanetary shock/ICME structures, *Journal of Geophysical Research: Space Physics*, 116, A12201, https://doi.org/10.1029/2011JA017013.

Zakharenkova, I., Cherniak, I., & Krankowski, A. (2021). Ground-Based GNSS and Satellite Observations of Auroral Ionospheric Irregularities during Geomagnetic Disturbances in August 2018, *Sensors*, 21(22), 7749. https://doi.org/10.3390/s21227749

Zhang, J., Dere, K. P., Howard, R. A., Kundu, M. R., & White, S. M. (2001). On the Temporal Relationship between Coronal Mass Ejections and Flares. *The Astrophysical Journal*, 559(1), 452–462. https://doi.org/10.1086/322405

Zhang, J., Richardson, I. G., Webb, D. F., Gopalswamy, N. Huttunen, E., Kasper, J.C., Nitta, N. V., Poomvises, W., et al. (2007). Solar and interplanetary sources of major geomagnetic storms (Dst < = 100 nT) during 1996–2005. *Journal of Geophysical Research: Space Physics*, 112(A10), A10102. https://doi.org/10.1029/2007JA012321

Zhao, L., Zurbuchen, T. H., Fisk, L. A. (2009). Global distribution of the solar wind during solar cycle 23: ACE observations. Geophysical Research Letters, Volume 36 (14), L14104. https://doi.org/10.1029/2009GL039181

Zhao, M. X., Le, G. M., Li, Q., Liu, G. A., & Mao, T. (2021). Dependence of Great Geomagnetic Storm ($\Delta$SYM-H$\leq$−200 nT) on Associated Solar Wind Parameters. *Solar Physics*, 296(4), 66. https://doi.org/10.1007/s11207-021-01816-2